\documentclass[aps,prl,twocolumn,longbibliography]{revtex4-2}
\usepackage{mathrsfs}
\usepackage{tabularx}
\usepackage{slashed}
\usepackage{amsmath}
\usepackage{amsfonts}
\usepackage{bm}
\usepackage{graphicx,color}
\usepackage{times}
\usepackage{braket}
\usepackage{orcidlink}
\usepackage[caption=false]{subfig}
\AtBeginDocument{%
   ~\newwrite\bibnotes
   ~\def\bibnotesext{NHFCI_1.bib}
   ~\immediate\openout\bibnotes=\jobname\bibnotesext
   ~\immediate\write\bibnotes{@CONTROL{REVTEX41Control}}
   ~\immediate\write\bibnotes{@CONTROL{%
    apsrev42Control,author="08",editor="1",pages="1",title="1",year="1"}}
    ~\if@filesw
    ~\immediate\write\@auxout{\string\citation{apsrev42Control}}%
   ~\fi
}%

\begin{document}


\title{Topological Order and Non-Hermitian Skin Effect in Generalized Ideal Chern Bands}

\author{Jiong-Hao Wang\orcidlink{0000-0002-4161-9614}}\thanks{jionghao.wang@fysik.su.se}
\author{Christopher Ekman\orcidlink{0009-0003-6426-2376}}
\author{Raul Perea-Causin\orcidlink{0000-0002-2229-0147}}
\author{Hui Liu\orcidlink{0009-0009-4988-9561}}
\author{Emil J. Bergholtz\orcidlink{0000-0002-9739-2930}}\thanks{emil.bergholtz@fysik.su.se}
\affiliation{Department of Physics, Stockholm University, SE-106 91 Stockholm, Sweden}

\date{\today}

\begin{abstract}
Fractionalization in ideal Chern bands and non-Hermitian topological physics are two active but so far separate research directions. Merging these, we generalize the notion of ideal Chern bands to the non-Hermitian realm and uncover several striking consequences both on the level of band theory and in the strongly interacting regime. 
Specifically, we show that the lowest band of a Kapit--Mueller lattice model with an imaginary gauge potential satisfies a generalized ideal condition with complex Berry curvature in sync with a complex quantum metric. 
The ideal band remains purely real and exactly flat on both the torus and cylinder: eigenstates are extended on the torus, while on the cylinder all right and left eigenstates localize at 
the boundaries, yielding a non-Hermitian skin effect without spectral winding. 
In the interacting regime, we find that the generalized ideal condition stabilizes an incompressible liquid at fractional fillings, retaining intrinsic non-Hermitian features on both cylinder and torus, while strikingly distinct on different manifolds.
On the cylinder, the ground states are always skin-Laughlin states.
In contrast, on the torus, we instead observe an unconventional competition between topologically ordered Laughlin-like states and negative collective modes, arising purely from non-Hermiticity.

\end{abstract}

\maketitle

Topological order is an organizing principle for quantum phases of strongly correlated systems with intrinsic long-range entanglement that lie beyond the conventional Landau paradigm of symmetry breaking~\cite{Wen2017RMP}. 
A famous example is the fractional quantum Hall effect~\cite{stormer1999nobel} and its zero-field analogue, fractional Chern insulators (FCIs)~\cite{Regnault2011PRX,LiuReview2024,bergholtz2013IJMP,PARAMESWARAN2013816,Kol1993PRB,Moller2009PRL,Kapit2010PRL,Tang2011PRL,Sun2011PRL,Nupert2011PRL,Sheng2011NC,Abouelkomsan2020PRL,Repellin2020PRR,Ledwith2020PRR,Liu2021PRL,Li2021PRR,Wilhelm2021PRB}, which exhibit quantized Hall conductance and support quasiparticles obeying fractional anyonic statistics.
Theoretical studies indicate that FCIs usually require topologically nontrivial flat bands with nearly ideal quantum geometry~\cite{Ledwidth2022PRL,WangJ2022PRL}, a condition recently realized experimentally in moiré materials, where zero-field FCIs have been observed at relatively high temperatures~\cite{Xie2021Nature,Park2023Nature,Cai2023Nature,Zeng2023Nature,Xu2023PRX,Lu2024Nature,Park2025arxiv}. 
These advances bring FCIs closer to practical applications in quantum devices, and may enable fault-tolerant quantum computing~\cite{nayak2008RMP}.
However, in realistic settings, unavoidable couplings to the environment, such as to leads or measurement apparatus, introduce gain and dissipation, making the system effectively non-Hermitian~\cite{Ashida2020AP}, which can significantly affect the stability and properties of topological orders. Specifically, dissipation is naturally present in platforms hosting FCIs of bosons in the form of cold atoms~\cite{Roncaglia2010PRL,Leonard2023Nature,Cooper2008}, photons \cite{Hafezi2013NJP,KapitPRX2014,Clark2020Nature,Kurilovich2022} and excitons \cite{XieExciton2024,PereaExciton2025}.

In parallel, non-Hermitian topology itself has recently emerged as a central topic of condensed matter, ranging from single-particle phenomena~\cite{Ashida2020AP,Bergholtz2021RMP,Gong2018PRX,Kawabata2019PRX,Song2019PRL,Yokomizo2019PRL,Ma2023PRB,li2021prb,wojcik2020prb,yang2024rpp} to many body effects~\cite{Lee2021PRB,Alsallom2022PRR,Yoshida2019SP,Lee2020PRB,Yang2021PRL,Kawabata2022PRB,Zhang2020PRB}, in which a variety of topological phases without Hermitian counterparts have been discovered~\cite{Bergholtz2021RMP,Ding2022NRP,Okuma2023ARCMP,Gong2018PRX,Kawabata2019PRX,Song2019PRL,Lee2020PRB,Kawabata2022PRB}.
Among these a striking manifestation is the so-called non-Hermitian skin effect (NHSE), where all eigenstates localize at system boundaries~\cite{Lee2016PRL,Xiong2018JPC,YaoWang2018PRL,Kunst2018PRL,Ma2023PRB,Martine2018PRB,Lee2021PRB,Alsallom2022PRR,LiuZ2024PRL,Gliozzi2024PRL,Shimomura2024PRL}, often linked to the topology of spectra under open boundary conditions (OBCs) and periodic boundary conditions (PBCs)~\cite{Okuma2020PRL,ZhangK2020PRL}, which has been observed in both classical systems~\cite{Brandenbourger2019NC,Xiao2020NatPhys,Helbig2020NatPhys,Weidemann2020Science} and open quantum systems~\cite{Liang2022PRL,Zhao2025Nature}.

While numerous studies have treated topological order and non-Hermitian physics separately, their cross-fertilization remains essentially unexplored. Although non-Hermitian approaches to FCIs have been theoretically explored in terms of dissipative preparation~\cite{Roncaglia2010PRL,Liu2021PRR} and particle loss~\cite{Hafezi2013NJP,Yoshida2019SP,Yoshida2020PRR}, these works arrive at Hermitian FCI states lacking the NHSE. 

Here, we show the interplay of topological order and non-Hermiticity. 
We consider the Kapit-Mueller (KM) model, a lattice realization of Landau levels supporting exactly flat topological bands. 
By introducing an imaginary gauge potential, we find that the lowest band remains exactly flat and purely real on both the torus and cylinder, while exhibiting a NHSE in the latter case, even without nontrivial spectral topology. 
The robustness of such a NHSE and the presence of the flat band on the cylinder are guaranteed by a similarity transformation between the Hermitian and non-Hermitian KM model.
On the torus, the failure of the similarity transformation manifests in the complex remote band, and the persistence of the flat band instead follows the intrinsic structure of the non-Hermitian eigenstates, which exhibits a complex generalization of ideal quantum geometry, dubbed as \textit{generalized ideal Chern band}. 

Switching on interactions then leads to the genuinely non-Hermitian FCIs, that combine features of incompressible topological liquids with non-Hermitian characteristics. 
On the cylinder, albeit the ground states remain exact zero modes, energetically analogous to Laughlin-like states, their particle density exhibits boundary accumulation, realizing a many-body skin effect. 
On the torus, the biorthogonal structure of non-Hermitian systems allows excited states to acquire negative energies, which compete with the nontrivial zero modes and drive an unconventional phase transition at a critical non-Hermiticity. 
Notably, this transition occurs much earlier than the single-particle gap-closing point, even within the single-band projection~\cite{yoshioka1983PRL}, indicating that it is an intrinsic property of the non-Hermitian ideal band and distinct from its Hermitian counterpart.
Consistently, the critical non-Hermiticity linearly grows with the interaction strength, reflecting a unique breakdown of the variational principle in non-Hermitian systems.    

\begin{figure}[!t]
  \centering
  \includegraphics[width=0.48\textwidth]{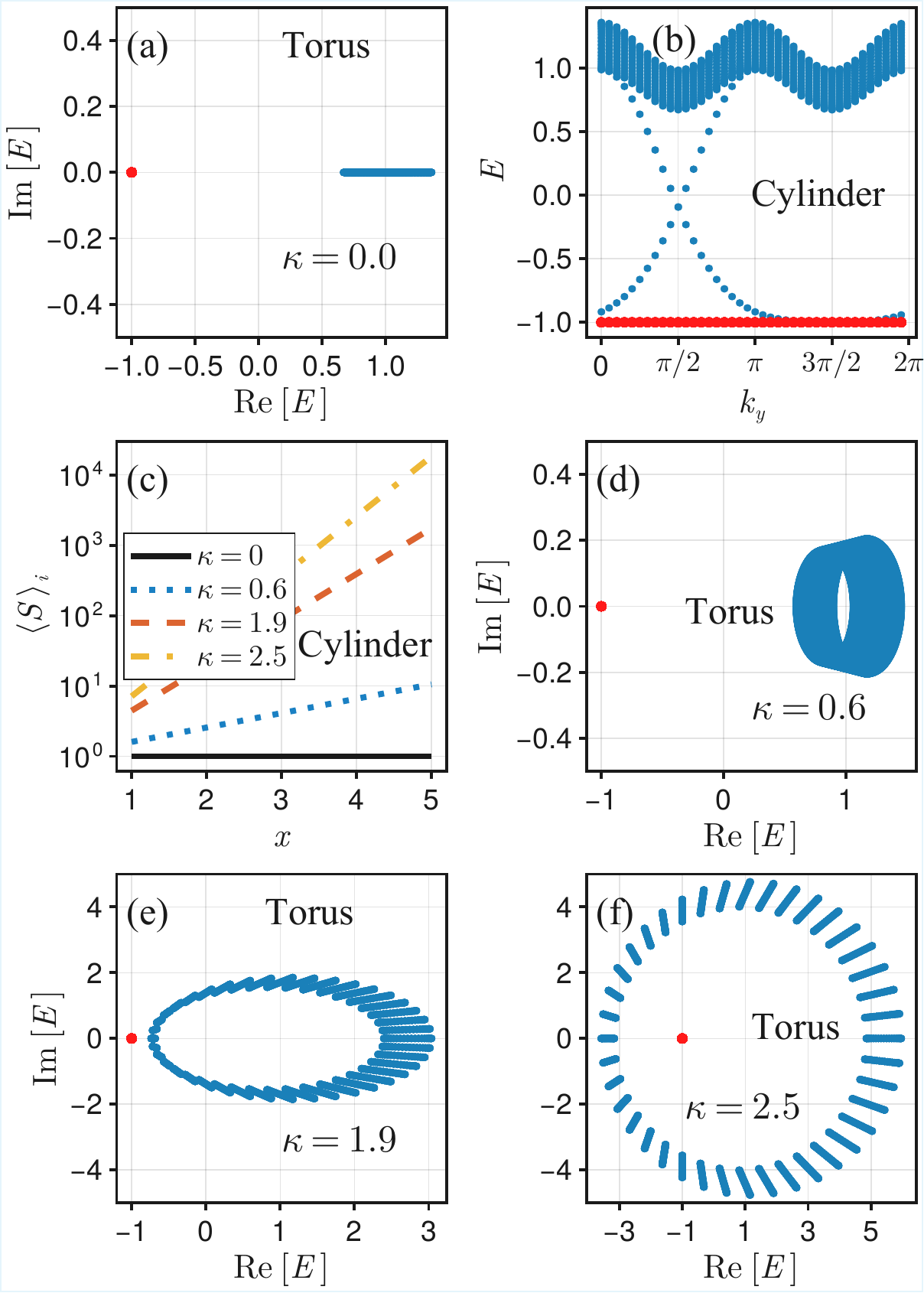}\hfill
  {\caption{\textbf{Non-Hermitian exact flat band}. (a) Energy spectrum of the Hermitian KM model on a torus. (b) Energy spectrum on the cylinder with OBCs in $x$, the same for Hermitian and non-Hermitian cases by virtue of the similarity transformation. Edge states emerge within the gap. (c) Skin-localization on the cylinder indicated by the expectation value $\langle S\rangle_i$ of the similarity transformation operator $S$ in terms of the single-particle state on lattice site $i$ as a function of the $x$ coordinate of site $i$. (d)-(f) Complex energy spectrum of the non-Hermitian KM model on the torus at $\kappa=0.6,1.9$ and $2.5$, respectively. For (a) and (d)-(f), the red dot represents the flat band with degeneracy equal to the number of magnetic fluxes, namely $N=L_xL_y/q$ on a square lattice with size $L_x,L_y$ and $\phi=1/q$. Here we take $q=2$.
  }
  \label{fig_single_particle}
  }
  
\end{figure}

\emph{Non-Hermitian Kapit-Mueller model.}--- We start from the tight-binding Hamiltonian with an imaginary gauge potential $\bm{A}=(i\kappa(\pi/2)(1-\phi),0,0)$:
\begin{equation}\label{NHKM}
H_0(\kappa)=\sum_{i\neq j}e^{-\kappa(\pi/2)(1-\phi)(x_j-x_i)}J_{ij}a_i^\dagger a_j,
\end{equation}
which leads to non-reciprocal hopping along the $x$-direction.
Here, $a^\dagger_i$ ($a_i$) creates (annihilates) a particle at site $(x_i,y_i)$ and $\kappa$ controls the strength of non-Hermiticity. 
The hopping amplitudes $J_{ij}$ are those of the original KM model~\cite{Kapit2010PRL} (see End matter for sketch and details). 
When $\kappa=0$, the system reduces to the original Hermitian case with rational magnetic flux $\phi$ through each plaquette in the Landau gauge, leading to an exactly flat band, as shown in Fig.~\ref{fig_single_particle}(a). 
Without loss of generality, we set $\phi=1/2$ throughout. 


On the cylinder with OBCs along $x$-direction, the Hamiltonian $H_0(0)$ and $H_0(\kappa)$ are related by a similarity transformation, $H_0(\kappa)=SH_0(0)S^{-1}$ with $S=e^{\kappa (\pi/2)(1-\phi)\sum_ix_ia^\dagger_ia_i}$ [see the proof in the Supplemental Material (SM)~\cite{SupMat}]. 
Consequently, both Hamiltonians share an identical energy spectrum, as shown in Fig.~\ref{fig_single_particle}(b), where edge states appear to connect the flat and conduction band.
However, the non-unitary $S$ matrix induces an asymmetric amplification and suppression along the $x$-direction, leading the corresponding left and right eigenstates to localize at opposite boundaries, forming a NHSE. 
This is illustrated in Fig.~\ref{fig_single_particle}(c), where the local expectation value $\langle S\rangle_i=\langle i|S|i\rangle$ with $|i\rangle=a_i^\dagger|0\rangle$ ($|0\rangle$ being the vacuum state) grows exponentially with $x$.

Closing the boundaries ill-defines such a similarity transformation. 
As a result, $H_0(\kappa)$ is no longer connected to its Hermitian counterpart, resulting in qualitatively different behavior on the torus. 
As shown in Fig.~\ref{fig_single_particle}(d-f), the remote band acquires complex energies as $\kappa$ increases, while the flat band remains exactly real and invariant, even after gap closing with the complex bands, as shown in Fig.~\ref{fig_single_particle}(f). 
This demonstrates that the presence of NHSE observed in the cylinder geometry is not tied to the well-known spectral winding, contrary to the usual expectation~\cite{Okuma2020PRL,ZhangK2020PRL}. 
The persistence of a purely real flat band on the torus follows from the intrinsic structure of the flat band, as detailed in the SM~\cite{SupMat}, instead of the similarity transformation on the cylinder.

\paragraph{Generalized ideal Chern band.---} 
To investigate the geometric and topological properties of the flat band, we calculate the non-Hermitian quantum geometric tensor defined as,
\begin{equation}\label{QGT}
G_{\mu\nu}=\langle\partial_{k_\mu} u^{L}_{\bm{k}}|\big(1-|u^{R}_{\bm{k}}\rangle\langle u^{L}_{\bm{k}}|\big )|\partial_{k_\nu} u^{R}_{\bm{k}}\rangle,
\end{equation}
where $\mu,\nu=x,y$ and $u^{L}_{\bm{k}}$ ($u^{R}_{\bm{k}}$) indicating the left/right eigenvector at momentum $\bm{k}=(k_x,k_y)$ of the flat band. 
The non-Hermitian quantum metric and the Berry curvature are obtained from $G_{\mu\nu}$ via $\text{Tr}(g)=\text{Tr}(G_{\mu\nu})$ and $\Omega_{xy}=i(G_{xy}-G_{yx})$, respectively.
This explicit use of both eigenvectors reflects the intrinsic biorthogonal structure, which is essential for defining geometric quantities, underlying the single-band projection in the interacting regime (see the next section and the SM~\cite{SupMat}). We also investigate quantum geometry under other definitions in the SM~\cite{SupMat}. 

\begin{figure}[!t]
  \centering
  \includegraphics[width=0.5\textwidth]{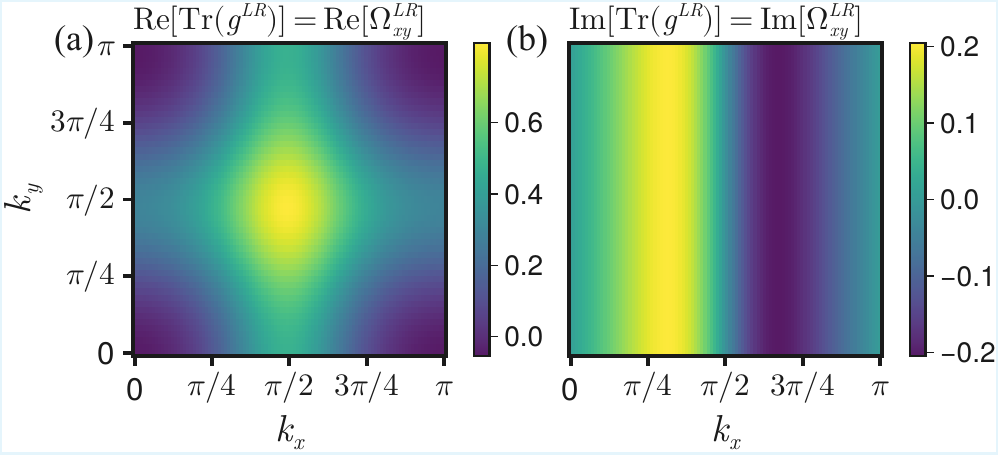}\hfill
  \caption{\textbf{Generalized ideal quantum geometry}. (a) Real and (b) imaginary part of $\mathrm{Tr}(g^{LR})$ and $\Omega_{xy}^{LR}$ which are exactly equal.
  For the gauge choice in Hamiltonian \eqref{NHKM} with flux $\phi=1/q$ ($q\in\mathbb{Z}$), the magnetic unit cell contains $q$ consecutive sites in the $x$ direction, so the magnetic Brillouin zone is $k_x\in [0,2\pi/q)$ and $k_y\in [0,2\pi)$.
  Owing to a residual translational symmetry by $\pi/q$ along $k_y$~\cite{Varjas2022SP}, we only display half of the Brillouin zone. 
  We take $\kappa=0.6$.
  }
  \label{fig_quant_geo}
\end{figure}

Due to the biorthogonal nature of non-Hermitian systems, we find the modified eigenstates now support a complex quantum geometry, as shown in Fig.~\ref{fig_quant_geo}(b), in contrast to the Hermitian case with real quantum geometry.
In this setting, the generalized ideal band condition takes the form 
\begin{align}
\mathrm{Re}[\mathrm{Tr}(g)]&=\pm\mathrm{Re}[\Omega_{xy}],\\
\mathrm{Im}[\mathrm{Tr}(g)]&=\pm\mathrm{Im}[\Omega_{xy}],
\end{align}
as shown in Fig.~\ref{fig_quant_geo} (see numerical confirmation in the SM~\cite{SupMat}),
which is a non-Hermitian generalization of the Hermitian ideal band condition~\cite{Ledwidth2022PRL,WangJ2022PRL}.
Despite the complexity, the integral of the complex Berry curvature over the Brillouin zone remains quantized, yielding an integer Chern number.  
Therefore, we term the flat band \textit{generalized ideal Chern band}, which may host topological orders exhibiting intrinsic non-Hermitian properties absent in their Hermitian counterparts.

\begin{figure}[t]
  \centering
  \includegraphics[width=0.48\textwidth]{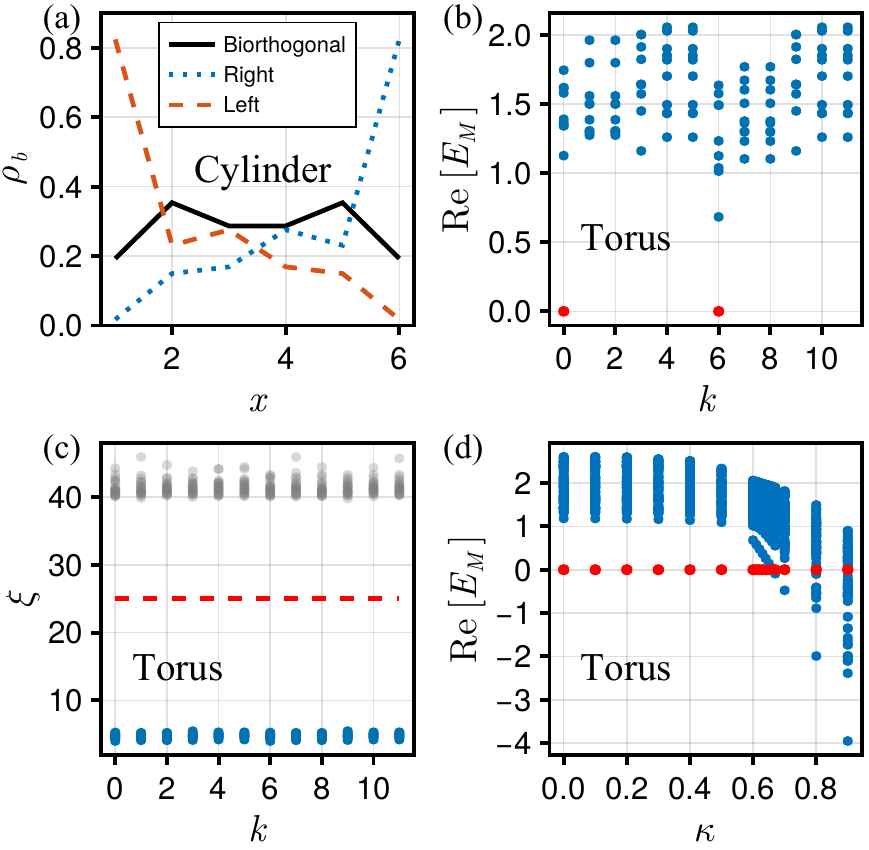}\hfill
  \caption{\textbf{Skin-Laughlin states}. (a) Particle number density $\rho_b$ of the ground state on the cylinder at $\kappa=1.9,U=0.1$ as a function of the $x$ coordinate for $L_x=6,L_y=3$ corresponding to total particle number $n_b=5$, 
  because of the integer shift $N=2n_b-1$ for Laughlin states on the cylinder~\cite{Liu2013PRB}, with $n_b$ being the particle number.
  The biorthogonal density, identical to the Hermitian case is represented by the black solid line, and the density of right and left eigenstates is represented by the blue dotted and red dashed line, respectively, showing the skin effect of the non-Hermitian Laughlin states.  
  (b) Real part of the many-body energy spectrum on the torus at $\kappa=0.6$ with the Laughlin states in red.
  (c) Particle entanglement spectrum of the Laughlin states in (b) for $n_A=3$.
  The number of states below the gap is 112, consistent with the generalized Pauli principle. 
  Note that $\xi$ above the gap should be infinite in principle, and the finite value here is due to machine precision.
  (d) Real part of the energy spectrum with respect to $\kappa$ with the Laughlin states in red, exhibiting the phase transition. 
  All results in (b)-(d) are obtained with system size $L_x=6,L_y=4$ at $\phi=1/2$ and filling factor $\nu=n_b/N=1/2$ for the band projected interaction,
  under which the interaction strength is the only energy scale, chosen arbitrarily here.
  }
  \label{fig_MBspec}
\end{figure}

\paragraph{Non-Hermitian topological order.---} We now explore interaction effects in a non-Hermitian ideal band. 
We consider a bosonic system with onsite repulsive interactions,
\begin{equation}\label{Int}
H(\kappa)=H_0(\kappa)+\sum_j \frac{U}{2}a^\dagger_ja^\dagger_ja_ja_j.
\end{equation}
In the Hermitian limit $\kappa=0$, the original KM model gives rise to Laughlin states at half filling~\cite{Kapit2010PRL}

On the cylinder with OBCs along $x$, the similarity transformation $H(\kappa)=SH(0)S^{-1}$ remains valid, which results in a many-body energy spectrum identical to that of the Hermitian case.
However, the corresponding Laughlin-like states are qualitatively modified: the zero modes accumulate near the boundaries, leading to the many-body skin effect. 
We demonstrate this by evaluating the ground state density $\rho_b(i)=\langle \Psi^X|a^{\dagger}_{i}a_{i}|\Psi^{X^\prime}\rangle$ as a function of the position along $x$-direction (where $X,X'\in{L, R}$ labeling the left or right eigenvectors). 
As shown in Fig.~\ref{fig_MBspec}(a),
the biorthogonal density constructed from both left and right eigenvectors remains uniform. 
In contrast, the densities with only left or right eigenvectors exhibit strong boundary localization, with particles accumulating at opposite edges, as hallmarks of skin-Laughlin modes. 

On the torus, as the similarity transformation fails, we numerically calculate the many-body energy spectrum by projecting the interaction onto the flat band~\cite{yoshioka1983PRL}. 
In this procedure, we employ the biorthogonal basis for consistency with the operator algebra (see SM~\cite{SupMat} for more details). 
In Fig.~\ref{fig_MBspec}(b), we show the real part of the energy at $\kappa=0.6$.
We observe two states at exactly zero energy [colored red in Fig.~\ref{fig_MBspec}(b)] with a finite gap to the excited states in the spectral flow (see SM~\cite{SupMat}).
The numerically obtained zero-energy states arise in the exact momentum sector consistent with the theoretical prediction of the generalized Pauli principle~\cite{Regnault2011PRX,Bernevig2008PRL}, as detailed in the End Matter, supporting FCIs. 
The energies of the zero modes are real while excited states can be complex (see SM~\cite{SupMat} for detailed discussion).
We remark that we view $H(\kappa)$ as an effective non-Hermitian Hamiltonian, as would arise from an energy-independent retarded self-energy. In that sense, the real parts of the eigenvalues define the many-body resonance energies (ground vs. excited sector), while the imaginary parts encode the corresponding lifetimes/broadenings.
We show that we can make the FCIs the longest-lived states by taking complex interaction $U=|U|e^{i\theta}$ arising from two-body loss in realistic setups~\cite{Yoshida2019SP} [see the SM~\cite{SupMat}].

To further rule out possible competing orders, we calculate the particle entanglement spectrum~\cite{Regnault2011PRX} [see the End Matter for details], as shown in Fig.~\ref{fig_MBspec}(c), where we take right eigenvectors.
The entanglement spectrum calculated by left eigenvectors is almost the same and that by biorthogonal eigenvectors has imaginary part but the real part is also very similar. 
We see a significant gap in the particle entanglement spectrum indicated by the red dashed line, and the number of states below the gap is consistent with the generalized Pauli principle~\cite{Regnault2011PRX}, confirming the zero modes as FCIs.
Beyond this numerical evidence, we further unveil the FCI character by analytically constructing the many-body wavefunction for the non-Hermitian Laughlin states on the torus, providing a direct and model-independent demonstration that complements the entanglement-spectrum analysis (as detailed in SM~\cite{SupMat}).

The robustness of the non-Hermitian Laughlin states against $\kappa$ on the torus is shown in Fig.~\ref{fig_MBspec}(d).
Here, the energy gap decreases with $\kappa$ and closes at $\kappa_c\sim 0.7$, which differs from the gap closing point of the single-particle spectrum at $\kappa_0\sim 2.0$.
After the gap closing, certain excitation states drop to negative energy after the gap closing point, in stark contrast to the Hermitian case where the energy is positive semi-definite for repulsive interactions, because minimizing kinetic and interaction energy separately do not necessarily minimize the real part of the total energy in the non-Hermitian regime [see SM~\cite{SupMat} for detailed explanation].
As $\kappa$ increases further, more and more states fall below zero energy, indicating a transition into a gapless regime as verified by the spectral flow [see SM~\cite{SupMat}].
Remarkably, there is always a pair of real-energy states pinned at zero energy even in the gapless regime, as indicated by red color in Fig.~\ref{fig_MBspec}(d), which are shown to be FCI states by the entanglement spectrum [see SM~\cite{SupMat}].
We also find the energies of the few states with the lowest real parts are always purely real and the states show signatures of superfluid order [see SM~\cite{SupMat}].

\begin{figure}[t]
  \centering
  \includegraphics[width=0.48\textwidth]{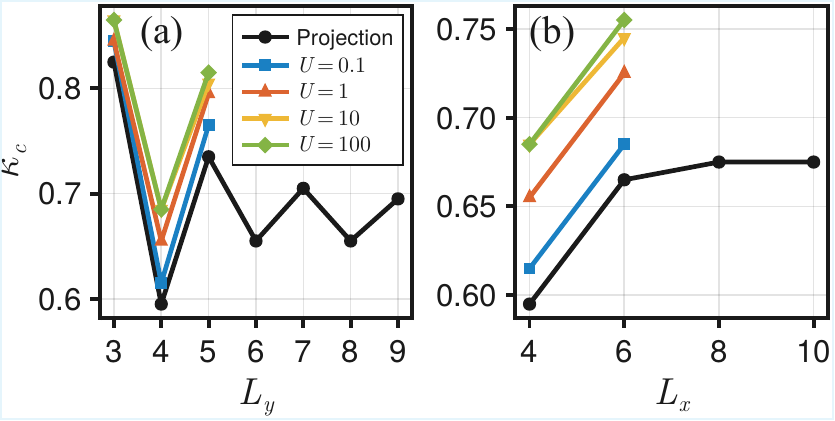}\hfill
  \caption{\textbf{Many-body transition point}. Critical $\kappa$ of the many-body phase transition with and without the single-band projection for various interaction strength for (a) $L_x=4$, with respect to $L_y$ and (b) $L_y=4$, with respect to $L_x$.
  }
  \label{fig_FiniteSize}
\end{figure}

We note that the presence of a finite $\kappa_c$ is not a finite-size effect. 
To show this, we calculate $\kappa_c$ for different system sizes. Our extrapolations converge for $\phi=1/2$ to $\kappa_c \sim 0.7$, well below the single-particle scale $\kappa_0\sim 2.0$. 
Due to computational restraints, we keep one of $L_x,L_y$ fixed and increase the other.
As shown by the black line in Fig.~\ref{fig_FiniteSize}(a), at $L_x=4$, $\kappa_c$ oscillates as $L_y$ increases for odd and even $L_y$ while the oscillating amplitude decays, showing a tendency to converge to a value about $0.7$.
For $L_y=4$, as shown in Fig.~\ref{fig_FiniteSize}(b),
$\kappa_c$ increases monotonically as $L_x$ only takes even values under our choice of magnetic unit cells and converges to $\kappa_c\sim 0.7$,
close to that in Fig.~\ref{fig_FiniteSize}(b), 
implying $\kappa_c$ remains finite and smaller than $\kappa_0$ in the thermodynamic limit.

Further including the remote band shows the persistence of such a many-body phase transition. 
Here, we evaluate the many-body energy spectrum without band projection as a function of interaction strength $U$, as shown in colored lines (compared to black lines) in Fig.~\ref{fig_FiniteSize}. 
All the way up to the hard-core limit, the transition point $\kappa$ is always smaller than $\kappa_0$. 
Based on intuition from the variational principle in Hermitian systems, one may expect that $\kappa_c$ cannot increase if we discard the single-band approximation, 
because the Hamiltonian is diagonalized within a larger Hilbert space.
Explicitly, according to the variational principle,
the enlarged Hilbert space leads to equal or lower ground state energy and thus a smaller $\kappa_c$ is expected to enter a gapless regime with negative energy.    
However, we find that $\kappa_c$ gets larger as we go beyond the single-band projection [see Fig.~\ref{fig_FiniteSize}], in stark contrast to the insight from Hermitian quantum mechanics,
which can be attributed to the breakdown of the conventional variational principle in non-Hermitian systems [see SM~\cite{SupMat} for a detailed explanation]. 

For generality, we also show results at $\phi=1/3$ in the SM~\cite{SupMat}, exhibiting the same physics qualitatively.
We have also constructed a Lindblad master equation within the framework of the dissipative open quantum system for the realization of the non-Hermitian FCIs with finite lifetime in the End matter.~\cite{SupMat,Torres2014PRA}.

\paragraph{Discussion.---}

We have investigated the interplay between topological order, the non-Hermitian skin effect, and generalized ideal quantum geometry at both the single-particle and many-body level. 
The non-interacting model hosts a purely real and exactly flat band which satisfies a generalized ideal condition for quantum geometry and furthermore exhibits the NHSE, thus providing an example outside the standard spectral-winding paradigm for the NHSE.
The physics of the interacting model depends on the underlying manifold. On a cylinder, the ground state is a skin-Laughlin state for all values of the non-Hermiticity $\kappa$. On a torus, in contrast, we found that Laughlin states are the ground states only for $\kappa$ below a critical value, while for larger $\kappa$, negative energy states outcompete the zero modes. This phase transition is attributed to intrinsic properties of the generalized ideal Chern band rather than to the influence of other bands. Going beyond the single-band projection, we found that the critical value of $\kappa$ increases with the interaction strength, demonstrating the breakdown of the variational principle familiar from Hermitian quantum mechanics.

A useful way to view our results is as a kind of competition between non-Hermitian skin effects and topological order, with a silver lining for subcritical $\kappa$ in that the lack of spectral winding allows for the survival of topological order. Despite the strong spectral sensitivity usually associated with the NHSE, for $\kappa<\kappa_c$, our numerics, along with general arguments for Laughlin-type topological order, indicate that the twofold ground-state manifold on the torus still realizes topological order in Wen's sense~\cite{Wen2017RMP}. The biorthogonal ground-state projector $P_{\rm GS} = \sum_{a=1}^2 |\Psi_a^R\rangle\langle\Psi_a^L|$ is adiabatically connected to that of the Hermitian Kapit--Mueller FCI, so the ground-state degeneracy, anyon structure and entanglement properties are expected to remain unchanged, and generic local perturbations that preserve the many-body gap should split the torus doublet only by amplitudes that are exponentially small in the system size. In this subcritical regime, non-Hermitian skin effects and topological order therefore genuinely coexist: the NHSE reshapes the right/left wave functions into ``skin-Laughlin'' states without destroying their topological content. For $\kappa>\kappa_c$ on the torus, however, the spectral rearrangement accompanied by the NHSE promotes more extended states to the bottom of the spectrum; the Laughlin manifold, while still present as biorthogonal eigenstates with the same topological signatures, moves out of the ground-state sector so that skin effect and topological order effectively compete.

Our work opens a new direction in the study of non-Hermitian topological order and indicates several intriguing avenues for future research.
First, the properties of our generalized ideal Chern bands suggest the existence of rigorous index theorems~\cite{DolanJPA2020} for many-body zero modes that apply more generally.
Second, the negative energy states on the torus require more detailed investigation by advanced numerical tools. 
Third, hidden connections between the many-body phase transition and intrinsic properties of the generalized ideal Chern band, like the complex quantum geometry, await to be uncovered.
Fourth, it is unclear whether the NHSE leads to novel responses of the FCI states.
Besides, it would be interesting to connect our findings to recent work on mixed-state topological order \cite{Fan2024PRXQ,Wang2025PRXQ,Ellison2025PRXQ} where non-unitary dynamics and coupling to environments are built-in from the outset.


Finally, while the main motivation of our work is conceptual we note that it may apply, mutatis mutandis, to systems ranging from AMO platforms such as rotating BECs~\cite{Cooper2008}, cold atoms~\cite{Leonard2023Nature,Cooper2019RMP} and photonics~\cite{Hafezi2013NJP,KapitPRX2014,Clark2020Nature,Kurilovich2022} where dissipation is ubiquitous, to solid state moiré heterostructures where bosonic FCIs are predicted to form in nearly ideal Chern bands of excitons with long but finite and tunable lifetimes~\cite{XieExciton2024,PereaExciton2025}. 
Specifically, although our calculations are based on the KM model, the results are generalizable to realistic continuous systems [see SM~\cite{SupMat}], which requires no fine tuning.
The required artificial gauge field~\cite{Lin2009Nature} and imaginary gauge potential~\cite{Tao2025arxiv} have been experimentally realized in continuous atomic gas, making it a promising platform.

\emph{Acknowledgments}---%
We thank helpful discussions with Kang Yang, Daniel Varjas, Fan Yang and Ai-Lei He.
This work was supported by the Swedish Research Council (2024-04567), the Knut and Alice Wallenberg Foundation (2023.0256), and the Göran Gustafsson Foundation for Research in Natural Sciences and Medicine.

\bibliography{NHFCI_1}

\section{End Matter}

\begin{figure}[h]
  \centering
  \includegraphics[width=0.3\textwidth]{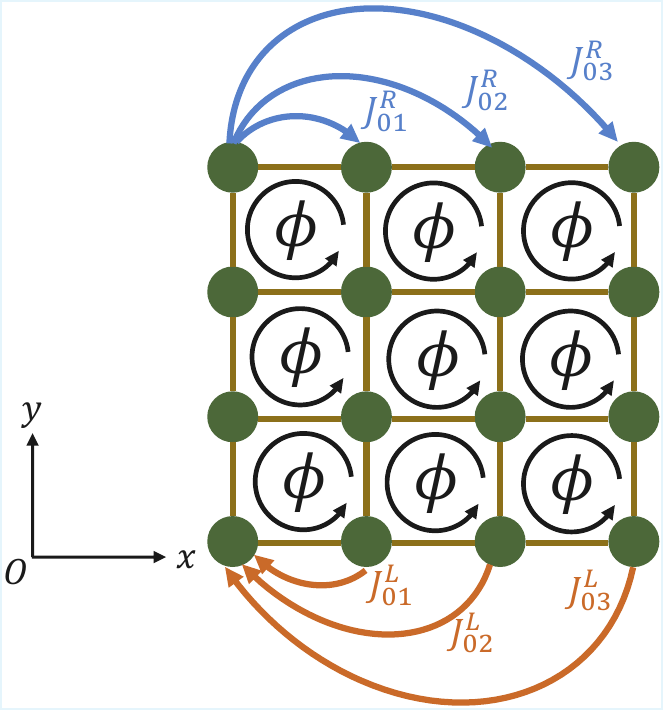}\hfill
  \caption{\textbf{Schematic of the non-Hermitian Kapit-Mueller model.} 
  The flux $\phi$ in each plaquette is represented by the counterclockwise arrowed circle. The hoppings to the right and left are indicated by blue and red colors, respectively, and the difference between the right and left hoppings leads to the non-Hermiticity.
  }
  \label{fig_sche}
        \end{figure}
\emph{Schematic of the non-Hermitian Kapit-Mueller model.---}
In this section, we show the schematic of the non-Hermitian KM model, described by Hamiltonian (\ref{NHKM}) in the main text, in Fig.~\ref{fig_sche}. 
 In Hamiltonian (\ref{NHKM}), $J_{ij}=W_{ij}e^{-i\pi y_{ij}(x_i+x_j)\phi}$ with $W_{ij}=(-1)^{ x_{ij}+ y_{ij}+x_{ij}y_{ij}}e^{-(\pi/2)(1-\phi)(x_{ij}^2+y_{ij}^2)}$, $x_{ij}=x_i-x_j$, and $y_{ij}=y_i-y_j$. 
Compared to Hofstadter model, the more famous lattice model with flux, the KM model is featured by the long-range hopping, with hoppings within the same row shown in Fig.~\ref{fig_sche} as examples.
The hopping from left to right $J_{0n}^R=e^{\kappa(\pi/2)(1-\phi)n}J_{ij}$ for $x_i-x_j=n,y_i=y_j$ (indicated by blue color) differs from that from right to left $J_{0n}^L=e^{-\kappa(\pi/2)(1-\phi)n}J^*_{ij}$ (indicated by red color) in strength, due to the exponential factor $e^{\pm\kappa(\pi/2)(1-\phi)n}$, leading to the non-Hermiticity. 

\emph{Generalized Pauli principle and particle entanglement spectrum.---}
The degeneracy of the FCIs and corresponding momentum sectors can be predicted by the generalized principle~\cite{Regnault2011PRX,Bernevig2008PRL}, which we explain here for $\nu=1/p=n_b/N$ bosonic Laughlin states with even $p$.
We rearrange the $N$ single-particle orbitals into a one-dimensional chain, and load $n_b$ particles into the orbitals. For $\nu=1/p$ bosonic Laughlin states, permissible configurations (called root configurations in Ref.~\onlinecite{Bernevig2008PRL}) satisfy the condition that, there exist at most one particle in every $p$ consecutive orbitals.
For $p=2$, there are two permissible configurations, $101010...$ and $010101...$, yielding two-fold degeneracy of FCIs, and the sums of the momenta of occupied orbitals in the configurations predict the associated momentum sectors.
If the degeneracy and momentum sectors obtained by numerics coincide with the predictions of the generalized Pauli principle, it suggests the existence of FCIs.  

For the particle entanglement spectrum, we partition the total $n_b$ bosons into $A$ and $B$ parts with $n_A$ and $n_B$ particles, respectively, and trace out the $B$ part, arriving at a reduced density matrix
$\rho_A=\mathrm{Tr}_B(\sum_i|\Psi^X_i\rangle\langle\Psi^{X^\prime}_i|/2)$
where the summation is over all the zero modes and $X=L/R$ represents left/right eigenvectors.
The eigenvalues of $\rho_A$ are $e^{-\xi}$ where $\xi$ is the particle entanglement spectrum.
There is a gap in the particle entanglement spectrum of FCIs, under which the number of states can also be predicted by the generalized Pauli principle for $\nu=1/p$ bosonic Laughlin states.
We load $n_A$ particles into the $N$ orbitals rearranged into one dimension, and the number of permissible configurations--at most one particle in every $p$ consecutive orbitals--is equivalent to the number of states under the gap in the particle entanglement spectrum.
The consistency between the numbers of states obtained numerically and theoretically is a further confirmation of FCIs.

\emph{Non-Hermitian FCI in open quantum systems.---}
In this section, we show how to realize the non-Hermitian FCIs in dissipative open quantum systems based on the Lindblad master equation and estimate the lifetime accordingly.

We consider an open quantum system described by the Lindblad master equation:
\begin{align}\label{Lindblad}
\frac{\mathrm{d}\rho}{\mathrm{d}t}&=-i[H_{\mathrm{KM}},\rho]+\sum_{i<j} (2L_{ij}\rho L_{ij}^\dagger-\{L_{ij}^\dagger L_{ij},\rho\})\\
&=-i(H_{\mathrm{eff}}\rho-\rho H_{\mathrm{eff}}^\dagger)+\sum_{i<j}2L_{ij}\rho L_{ij}^\dagger,
\end{align}
where $H_{\mathrm{KM}}=\sum_{i\neq j}J_{ij}a^\dagger_ia_j$ is the Hermitian Hamiltonian of the pristine KM model with $J_{ij}$ defined in the main text, the Lindblad operator $L_{ij}=A_{ij}a_i-iB_{ij}a_j$ characterizes the dissipation, and the effective non-Hermitian Hamiltonian is
$H_{\mathrm{eff}}=H_{\mathrm{KM}}-i\sum_{i<j}L_{ij}^{\dagger}L_{ij}$.
As $L_{ij}^{\dagger}L_{ij}=|A_{ij}|^{2}a_{i}^{\dagger}a_{i}+|B_{ij}|^{2}a_{j}^{\dagger}a_{j}+iB_{ij}^{*}A_{ij}a_{j}^{\dagger}a_{i}-iA_{ij}^{*}B_{ij}a_{i}^{\dagger}a_{j}$,
if we choose $A_{ij}^{*}B_{ij}=\gamma(x_{i}-x_{j})J_{ij}$,
we have 
\begin{align}
H_{\mathrm{eff}}=\sum_{i<j}J_{ij}[1-\gamma(x_j-x_i)]a_i^\dagger a_j\nonumber\\
+J^*_{ij}[1-\gamma(x_i-x_j)]a_j^\dagger a_i-i\Delta_i a_i^\dagger a_i,
\end{align}
with $\Delta_i=\sum_{i<j}|A_{ij}|^2+\sum_{j<i}|B_{ji}|^2$.
For $\gamma=\kappa(\pi/2)(1-\phi)$, $H_{\mathrm{eff}}$ is equivalent to the non-Hermitian KM model Eq.~\ref{NHKM} in the main text to the first order of $\gamma$ except for the loss term $-i\Delta_i a_i^\dagger a_i$. 
By setting $A_{ij}=\sqrt{\gamma|(x_i-x_j)J_{ij}|},B_{ij}=\gamma(x_{i}-x_{j})J_{ij}/\sqrt{\gamma|(x_i-x_j)J_{ij}|}$,
we see $\Delta_i$ is constant on a torus due to the translational symmetry, and approximately constant in the bulk under OBCs as $\Delta_i$ is dominated by the nearest few sites around site $i$ due to the exponential decay of $J_{ij}$.
Therefore, the effect of the loss term $-i\Delta_i a_i^\dagger a_i$ in $H_{\mathrm{eff}}$ can be accounted for as a constant shift of each eigenenergy by an imaginary value $-i\Delta n_b$ with $\Delta=\Delta_i$ and $n_b$ being the number of particles. 

For open quantum systems, we concern about the eigensystem of the Liouvillian operator $\mathcal{L}$ in the Lindblad master equation $\mathrm{d}\rho/\mathrm{d}t=\mathcal{L}\rho$,
that is, $\mathcal{L}\rho=\lambda \rho$, with $\mathrm{Re}[\lambda]<0$ corresponding to the dissipative case associated with lifetime $\tau\sim 1/\mathrm{Re}[-\lambda]$.
The explicit form of $\mathcal{L}$ can be found in Eq.~\ref{Lindblad}.
It is proved~\cite{Torres2014PRA} that in the presence of loss only or gain only, the eigenstate is 
\begin{align}
\rho_{jk}=|\psi^R_j\rangle \langle \psi^R_k|
\end{align}
with eigenvalue $\lambda_{jk}=-i(E_j-E^*_k)=\mathrm{Im}(E_j)+\mathrm{Im}(E_k)$, where $|\psi^R_j\rangle$ is the right eigenvector of $H_{\mathrm{eff}}$ with eigenenergy $E_j$. 
For the pure state of non-Hermitian Laughlin states, i.e., $\rho_{\mathrm{Lau}}=|\psi^R_{\mathrm{Lau}}\rangle \langle \psi^R_{\mathrm{Lau}}|$, the corresponding eigenvalue is $\lambda_{\mathrm{Lau}}=2\mathrm{Im}(E_{\mathrm{Lau}})$.
In the main text, the energy of the non-Hermitian Laughlin state is purely real and exactly zero. 
For the effective Hamiltonian arising from the dissipative Lindblad master equation, due to the extra loss term $E_{\mathrm{Lau}}=-i\Delta n_b$, the eigenvalue of the Liouvillian operator is $\lambda_{\mathrm{Lau}}=-2\Delta n_b$, corresponding to the lifetime $\tau \sim 1/(2\Delta n_b)$. 
Note that due to the loss, the steady state is the vacuum.

In the SM~\cite{SupMat}, we have shown that by taking a complex interaction $U=|U|e^{i\theta}$, we can render the non-Hermitian FCIs the only states with zero imaginary part of the energy while all the other states have nonzero negative imaginary part.
The imaginary interaction can be realized by two-body loss, i.e., the Lindblad operator $L_i\sim a_ia_i$~\cite{Yoshida2019SP}.
In this case, taking into account the imaginary energy shift $i\Delta n_b$, the energy of FCIs has the smallest absolute value of the imaginary part.
Subsequently, the Laughlin state $\rho_{\mathrm{Lau}}$ has the smallest absolute value of the imaginary part of the Liouvillian eigenvalue $\lambda$, corresponding to the longest lifetime in the specific particle number sector.

\onecolumngrid

\section{Supplemental Material}


	\setcounter{equation}{0} \setcounter{figure}{0} \setcounter{table}{0} %
	\renewcommand{\theequation}{S\arabic{equation}} \renewcommand{\thefigure}{S%
		\arabic{figure}}

In the Supplemental Material, we show numerical evidence of the equality between complex quantum metric and Berry curvature, 
prove the similarity transformation for relating Hermitian and non-Hermitian Hamiltonian on a cylinder, show analytical wave functions of both single-particle states on the flat band and non-Hermitian Laughlin states, 
demonstrate the quantum geometry calculated with both left and both right eigenvectors with analytical explanations, 
show additional details on the single-band projection procedure in the non-Hermitian context and numerical proof, 
illustrate the emergence of negative energy states and breakdown of variational principle in non-Hermitian systems, 
discuss the stability of the non-Hermitian FCIs within the framework of the Lindblad master equation, and provide more numerical data on spectral flow, entanglement spectrum, signatures of superfluid, many-body energy spectrum beyond the single-band projection and results for $\phi=1/3$. 

\section{Numerical confirmation of ideal quantum geometry}

\begin{figure}[h]
  \centering
  \includegraphics[width=0.6\textwidth]{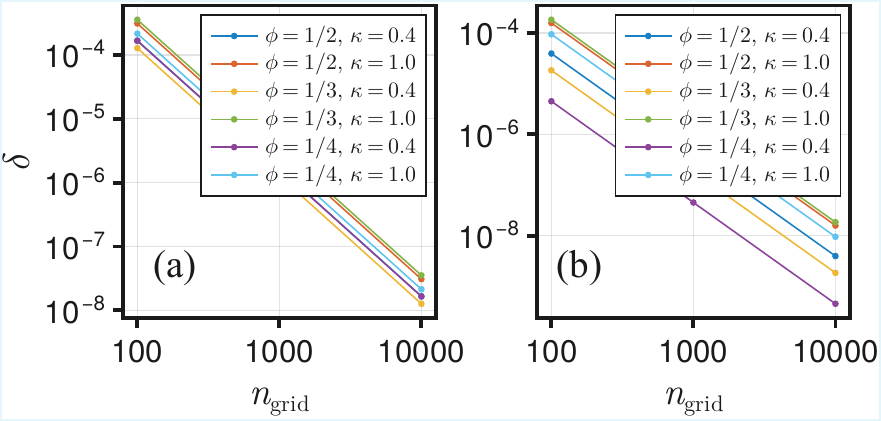}\hfill
  \caption{\textbf{}Difference of the (a)Real and (b) imaginary part between quantum metric $g^{LR}$ and Berry curvature $\Omega_{xy}^{LR}$ versus the grid density $n_{\mathrm{grid}}$.
  }
  \label{Fig_Geo_Diff}
        \end{figure}

In this section, we justify the ideal quantum geometry numerically by showing the real and  imaginary part of $\delta=\mathrm{Tr}(g^{LR})-\Omega_{xy}^{LR}$ in Fig.~\ref{Fig_Geo_Diff}(a) and (b) with respect to the grid density  $n_{\mathrm{grid}}$, i.e., calculated under a $n_{\mathrm{grid}}\times n_{\mathrm{grid}} $ grids in the momentum space. 
We see $\delta$ decreases and tends to zero as $n_{\mathrm{grid}}$ increases, indicating the ideal conditions (7) and (8) in the main text are satisfied.

\section{Similarity transformation for non-Hermitian skin effect}
In this section, we prove both the single-particle and many-body Hamiltonian under non-Hermitian deformation in the main text can be related to their Hermitian counterparts by a similarity transformation on a cylinder, from which the non-Hermitian skin effect is clear. We take the second quantization formulation in the proof.

The single-particle Hamiltonian of the Kapit-Mueller (KM) model in Eq.1 of the main text reads
\begin{align}
H_0(\kappa)=\sum_{i\neq j}e^{-\gamma(x_j-x_i)}J_{ij}a_i^\dagger a_j,
\end{align}
where we denote $\gamma=\kappa(\pi/2)(1-\phi)$ for simplicity.
The similarity transformation operator is $S(\kappa)=e^{\gamma\sum_ix_ia^\dagger_ia_i}=e^{\gamma\sum_ix_in_i}$ with $n_j=a^\dagger_ja_j$ being the particle number operator.
As the number operators commute with each other, we have  
$S^{-1}(\kappa)=e^{-\gamma\sum_ix_in_i}$.
In the following, we prove the relation 
\begin{align}\label{SingleSim}
H_0(\kappa)=S(\kappa)H_0(0)S^{-1}(\kappa)
\end{align}
independent of whether $a^\dagger_i$ ($a_i$) is the creation (annihilation) operator of bosons or fermions.
The most important tool in the proof is the Baker-Campbell-Hausdorff (BCH) formula
\begin{align}
e^ABe^{-A}=B+[A,B]+\frac{1}{2!}[A,[A,B]]+...+\frac{1}{n!}[A,[A,...[A,B]]]
\end{align}
where $A,B$ are operators, $[...]$ denotes the commutator and $n$ is an integer tending to infinity.
Using the BCH formula, we can directly obtain
\begin{align}
S(\kappa)a_i^\dagger a_jS^{-1}(\kappa)&=a_i^\dagger a_j+\gamma(x_i-x_j)a_i^\dagger a_j+\frac{1}{2!}[\gamma(x_i-x_j)]^2a_i^\dagger a_j+...+\frac{1}{n!}[\gamma(x_i-x_j)]^na_i^\dagger a_j\\
&=e^{\gamma(x_i-x_j)}a_i^\dagger a_j,
\end{align}
where we have used the equation $[\sum_mx_ma^\dagger_ma_m,a_i^\dagger a_j]=(x_i-x_j)a_i^\dagger a_j$, which can be proved easily for both bosons and fermions.
Therefore, the single-particle case Eq.~\ref{SingleSim} is proved.

The many-body Hamiltonian for interacting bosons in Eq.~9 in the main text is $H(\kappa)=H_0(\kappa)+H_I$ with
\begin{align}
H_I=\sum_j \frac{U}{2}a^\dagger_ja^\dagger_ja_ja_j=\sum_j \frac{U}{2}n_j(n_j-1)
\end{align}
As number operators commute, it is obvious that $H_I$ is invariant under the similarity transformation by $S$ also from BCH formula.
Combining Eq.~\ref{SingleSim}, we arrive at $H(\kappa)=S(\kappa)H(0)S^{-1}(\kappa)$ for the total Hamiltonian of interacting bosons.
Therefore, the right and left eigenvectors $|\Psi^R(\kappa)\rangle$ and $|\Psi^L(\kappa)\rangle$ of $H(\kappa)$ are related to the eigenstate $|\Psi\rangle$ of the Hermitian Hamiltonian $H(0)$ by
\begin{align}
|\Psi^R(\kappa)\rangle&=S(\kappa)|\Psi\rangle,\\
\langle \Psi^L(\kappa)|&=\langle \Psi|S^{-1}(\kappa),
\end{align}
leading to the non-Hermitian skin effect.

By the similarity transformation, we can prove on the cylinder the biorthogonal number density defined in the main text 
\begin{align}
\rho_b(i)&=\langle \Psi^L(\kappa)|a^{\dagger}_{i}a_{i}|\Psi^R(\kappa)\rangle\\
&=\langle \Psi|S^{-1}(\kappa)a^{\dagger}_{i}a_{i}S(\kappa)|\Psi\rangle\\
&=\langle \Psi|a^{\dagger}_{i}a_{i}|\Psi\rangle,
\end{align}
which is identical to the particle number density of the Hermitian state $|\Psi\rangle$.
In the proof, we have used the commutating relation $[S(\kappa),a^{\dagger}_{i}a_{i}]=0$.

\section{Analytical wave functions on the torus}\label{AWF}
In this section, we give analytical wave functions on the torus for both the flat band of the non-Hermitian KM model at the single-particle level and the non-Hermitian FCI states in the interacting regime.

First, we review the standard textbook problem of a charged particle under a magnetic field with strength $B$ corresponding to the vector potential ${\bm A}=(A_x,A_y,0)$  in the continuum with Hamiltonian
\begin{align}
H_{con}=\frac{1}{2m}[\pi_x^2+\pi_y^2],
\end{align}
where $q,m$ are electric charge and mass of the particle, respectively, and $\pi_\mu=-i\hbar\partial_\mu-qA_\mu$ with $\mu=x,y$ and $\hbar$ being the Planck constant.
By defining ladder operators 
\begin{align}
a&=\frac{l_{B}}{\sqrt{2}\hbar}\bigg(i\pi_{x}-\pi_{y}\bigg),\\
a^{\dagger}&=\frac{l_{B}}{\sqrt{2}\hbar}\bigg(-i\pi_{x}-\pi_{y}\bigg),
\end{align}
with magnetic length $l_B=\sqrt{\hbar/(qB)}$,
we arrive at 
\begin{align}\label{HamLL}
H_{con}=\hbar \omega (a^\dagger a+\frac{1}{2})
\end{align}
with $\omega=qB/m$.
As eigenvalues of $a^{\dagger}a$ take non-negative integer values, the Hamiltonian (\ref{HamLL}) yields Landau level energy spectrum $E_{LL}=\hbar\omega/2,3\hbar\omega/2,5\hbar\omega/2,...$.
The wave functions on the lowest Landau level satisfy $a\psi(x,y)=0$.
If we choose the symmetric gauge $\bm{A}=(By/2,-Bx/2,0)$ for the convenience in constructing Laughlin wave functions in the following, we have 
\begin{align}
a&=\frac{1}{\sqrt{2}}\bigg(2l_{B}\partial_{\bar{z}}+\frac{z}{2l_{B}}\bigg)\\
a^\dagger&=\frac{1}{\sqrt{2}}\bigg(-2l_{B}\partial_{z}+\frac{\bar{z}}{2l_{B}}\bigg).
\end{align}
For a system with lengths $L_1,L_2$ along two directions separated by an angle $\theta$, i.e., $\bm{L}_1=(L_1,0),\bm{L}_2=(L_2\cos\theta,L_2\sin\theta)$ the wave function under PBCs has the form~\cite{Haldane1985PRB}
\begin{align}
\psi(z)=e^{(z^2-|z|^2)/(4l_B^2)}\prod_{\nu=1}^{N_\phi}\theta_1(\pi(z-z_\nu)/L_1|\tau)
\end{align}
where the odd elliptic theta function $\theta_1(z|\tau)=\sum_{n=-\infty}^{\infty}(-1)^{n+1/2}e^{i\pi\tau(n+1/2)^{2}}e^{i2z(n+1/2)}$ with $\tau=e^{i\theta}L_2/L_1$, $N_\phi$ is the degeneracy of the Landau level and $z_\nu$ are parameters adequately chosen to satisfy the specific boundary conditions under magnetic translations.
Explicitly,
\begin{align}
\psi(z+L_1)&=(-1)^{N_{\phi}}\exp\bigg(iL_1y/(2l_B^2)\bigg)\psi(z),\label{BCx}\\
\psi(z+e^{i\theta}L_2)&=(-1)^{N_{\phi}}\exp\bigg(-L_2(e^{i\theta}z^{*}-e^{-i\theta}z)/(2l_B^2)\bigg)\exp\bigg[i2\pi\frac{\sum_{v}z_{\nu}}{L_1}\bigg]\psi(z).\label{BCy}
\end{align}
If $N_\phi$ is even, the magneto-periodic boundary conditions constrain $\sum_{v}z_{\nu}=nL_1$ with integer $n$.

By introducing the imaginary gauge potential $\bm{A}_{imag}=(i\gamma,0,0)$, we obtain the non-Hermitian Hamiltonian 
\begin{align}
H^{NH}_{con}=\frac{1}{2m}[(\pi_x-iq\gamma)^2+\pi_y^2],
\end{align}
which can be rewritten as 
\begin{align}\label{HamNH}
H^{NH}_{con}=\hbar\omega(b^+b+\frac{1}{2}),
\end{align}
with 
\begin{align}
b&=a+\frac{el_{B}}{\sqrt{2}\hbar}\gamma, \label{b1}\\
b^{+}&=a^{\dagger}-\frac{el_{B}}{\sqrt{2}\hbar}\gamma,\label{b2}
\end{align}
where we use $+$ instead of $\dagger$ to indicate $b^+$ and $b$ are not Hermitian conjugate of each other.
The derivation of (\ref{HamNH}) is independent of the boundary conditions, therefore the Landau level spectrum does not change under the non-Hermitian deformation even on the torus.
Under the symmetric gauge, it is easy to see 
\begin{align}
b&=\frac{1}{\sqrt{2}}\bigg(2l_{B}\partial_{\bar{z}^R}+\frac{z^R}{2l_{B}}\bigg),\\
b^+&=\frac{1}{\sqrt{2}}\bigg(-2l_{B}\partial_{{z}^L}+\frac{\bar{z}^L}{2l_{B}}\bigg)
\end{align}
with $z^R=z+2ql_B^2\gamma/\hbar,z^L=z-2ql_B^2\gamma/\hbar$.
Due to $b\psi^R(z)=0$ and $(b^+)^\dagger\psi^L(z)=0$, we arrive at the right and left eigenvectors under the imaginary gauge potential:
\begin{align}
\psi^R(z)=e^{((z^R)^2-|z^R|^2)/(4l_B^2)}\prod_{\nu=1}^{N_\phi}\theta_1(\pi(z^R-\tilde{z}^R_\nu)/L_1|\tau),\\
\psi^L(z)=e^{((z^L)^2-|z^L|^2)/(4l_B^2)}\prod_{\nu=1}^{N_\phi}\theta_1(\pi(z^L-\tilde{z}^L_\nu)/L_1|\tau),
\end{align}
which are opposite spatial shifts along $x$ direction.  
It is easy to see $\psi^X(z)$ with $X=R,L$ satisfies Eq.~\ref{BCx} while an extra phase factor $e^{\mp iq\gamma L_y/\hbar}$ arises, so to preserve the boundary conditions, we have $\sum_{\nu}\tilde{z}^R_{\nu}-\sum_{\nu}z_{\nu}=q\gamma L_1L_2/(2\pi\hbar)$ and $\sum_{\nu}\tilde{z}^L_{\nu}-\sum_{\nu}z_{\nu}=-q\gamma L_1L_2/(2\pi\hbar)$.

On the original KM model, the wave function on the flat band is the lattice discretization of the lowest Landau level wave function in the continuum~\cite{Kapit2010PRL} by substituting $1/l_B^2$ with $2\pi \phi$.
It is also the case under the imaginary gauge potential, i.e., the wave functions read
\begin{align}
\psi^R_{KM}(z)=e^{((z^R)^2-|z^R|^2)\pi\phi/2}\prod_{\nu=1}^{N_\phi}\theta_1(\pi(z^R-\tilde{z}^R_\nu)/L_1|\tau),\label{REV}\\
\psi^L_{KM}(z)=e^{((z^L)^2-|z^L|^2)\pi\phi/2}\prod_{\nu=1}^{N_\phi}\theta_1(\pi(z^L-\tilde{z}^L_\nu)/L_1|\tau),\label{LEV}
\end{align}
on the non-Hermitian KM model, 
where $z^R=z+\gamma/(\pi\phi),z^L=z-\gamma/(\pi\phi)$ corresponding to $\bm{A}_{imag}=(i\gamma,0,0)$,
and $\sum_{\nu}\tilde{z}^R_{\nu}-\sum_{\nu}z_{\nu}=L_1L_2\gamma/(2\pi),\sum_{\nu}\tilde{z}^L_{\nu}-\sum_{\nu}z_{\nu}=-L_1L_2\gamma/(2\pi)$ to keep the boundary conditions intact.
In the following, we will prove mathematically.

We reproduce the Hamiltonian of the non-Hermitian KM model on a torus here:
\begin{align}
H_0(\gamma)=\sum^{z_j\neq z_k+R}_{j,k,R}\Gamma(z_j,z_k+R,\gamma)J(z_j,z_k+R)P(z_k,R)a^\dagger_ja_k,
\end{align}
where $R=nL_x+me^{i\theta}L_y$ with integer $n,m$.
For the lattice model, we consider the case with $\theta=\pi/2$ for simplicity, so we use $L_x,L_y$ instead of $L_1,L_2$ afterwards.
Here, $\Gamma(z_j,z_k,\gamma)=e^{\gamma (z_{kj}+z_{kj}^*)/2}$ with $z_{kj}=z_k-z_j$ being the non-reciprocal hopping, $J(z_j,z_k)=(-1)^{x_{jk}+y_{jk}+x_{jk}y_{jk}}e^{-(\pi/2)[(1-\phi)|z|^{2}]}e^{(\pi/2)(z_{j}z^{*}-z_{j}^{*}z)\phi}$ with $x_{kj}=x_k-x_j,y_{kj}=y_k-y_j$ is the hopping term of the pristine KM model, and $P(z_k,R)=e^{\pi\phi(z_{k}R^{*}-z_{k}^{*}R)/2}$ is the extra phase for hopping across the boundaries, where $z=x+iy$ and $u_{kj}=u_k-u_j$ for $u=z,x,y$.
From Eq.~\ref{BCx} and \ref{BCy}, we have
\begin{align}
\psi^R(z+R)=P(z,R)\psi^R(z)
\end{align}
if we choose $N_\phi$ is even and $\sum_\nu z_\nu$ is an integer multiple of $L_x$.
Then we have
\begin{align}
\frac{\langle j|H_0(\gamma)|\psi^{R}\rangle}{\langle j|\psi^{R}\rangle}&=\frac{\sum^{z_j\neq z_k+R}_{k,R}\Gamma(z_j,z_k+R,\gamma)J(z_j,z_k+R)P(z_k,R)\psi^R(z_k+\frac{\gamma}{\pi\phi})}{\psi^R(z_j)}\\
&=\frac{\sum^{z_j\neq z_k+R}_{k,R}\Gamma(z_j,z_k+R,\gamma)J(z_j,z_k+R)P(z_k,R)P^{-1}(z_k,R)\psi^R(z_k+\frac{\gamma}{\pi\phi}+R)}{\psi^R(z_j)}\\
&=\sum^{z_j\neq z_k+R}_{k,R}\prod_{\nu=1}^{N_{\phi}}\frac{\exp\bigg(\frac{\pi\phi}{2}(z_{j}+z_{jk}+\frac{\gamma}{\pi\phi}+R)^{2}\bigg)\theta_{1}[\pi(z_{j}+z_{jk}+\frac{\gamma}{\pi\phi}+R-\tilde{z}^R_{\nu})/L_x|\tau]}{\exp\bigg(\frac{\pi\phi}{2}\big(z_{j}+\frac{\gamma}{\pi\phi}\big)^{2}\bigg)\theta_{1}[\pi(z_{j}+\frac{\gamma}{\pi\phi}-\tilde{z}^R_{\nu})/L_x|\tau]}\nonumber\\
&\exp\bigg(-\pi\phi z_{j}^{*}(z_{jk}+R)\bigg)G(z_{jk}+R)\exp\bigg(-\frac{\pi}{2}|z_{jk}+R|^{2}\bigg)\label{Eminus1}.
\end{align}
Due to the relation $\sum_zf(z)G(z)e^{-(\pi/2)|z|^2}=0$ for any entire function $f(z)$, we have
$\langle j|H_0(\gamma)|\psi^{R}\rangle/\langle j|\psi^{R}\rangle=-1$ because the summation in Eq.~\ref{Eminus1} discards $z_{jk}+R=0$. 
Similarly, we can also prove $\langle j|H_0^\dagger(\gamma)|\psi^{L}\rangle/\langle j|\psi^{L}\rangle=-1$.
Therefore, we have proved Eq.~\ref{REV} and \ref{LEV} are eigen wave functions of the non-Hermitian KM model at energy $-1$, which also explains the persistence of the flat band under non-Hermitian deformation.

The Laughlin wave function at filling factor $\nu=1/p$ for the Hermitian KM model on a torus is~\cite{Kapit2010PRL}
\begin{align}
\Psi({z_n})=\prod_{i=1}^p\theta_1\bigg(\frac{\pi}{L}(Z-Z_i)\bigg|\tau\bigg)\prod_{k<j}^{N}\theta_1\bigg(\frac{\pi}{L}(z_j-z_k)\bigg|\tau\bigg)^p\prod_{j=1}^N e^{(z_j^2-|z_j|^2)\pi\phi/2},
\end{align}
where $N$ is the number of particles, $Z=\sum_jz_j$ is the center of mass coordinate and $Z_i$ are parameters.
From Eq.(\ref{REV}) and (\ref{LEV}), we deduce the right and left eigenvectors of the zero modes on the non-Hermitian KM model under onsite interaction are 
\begin{align}
\Psi^R({z_n},\kappa)&=\prod_{i=1}^p\theta_1\bigg(\frac{\pi}{L}(Z^R-\tilde{Z}^{R}_i\bigg|\tau\bigg)\prod_{k<j}^{N}\theta_1\bigg(\frac{\pi}{L}(z_j-z_k)\bigg|\tau\bigg)^p\prod_{j=1}^N e^{((z_j^R)^2-|z_j^R|^2)\pi\phi/2},\\
\Psi^L({z_n},\kappa)&=\prod_{i=1}^p\theta_1\bigg(\frac{\pi}{L}(Z^L-\tilde{Z}^{L}_i)\bigg|\tau\bigg)\prod_{k<j}^{N}\theta_1\bigg(\frac{\pi}{L}(z_j-z_k)\bigg|\tau\bigg)^p\prod_{j=1}^N e^{((z_j^L)^2-|z_j^L|^2)\pi\phi/2}
\end{align}
with $\sum_{i}\tilde{Z}^R_{i}-\sum_{i}Z_{i}=L_xL_y\gamma/(2\pi),\sum_{i}\tilde{Z}^L_{i}-\sum_{i}Z_{i}=-L_xL_y\gamma/(2\pi)$ to satisfy the original boundary conditions.
which we have numerically confirmed. 

\section{Quantum geometry calculated with right-right or left-left eigenvectors}

\begin{figure}[h]
  \centering
  \includegraphics[width=0.8\textwidth]{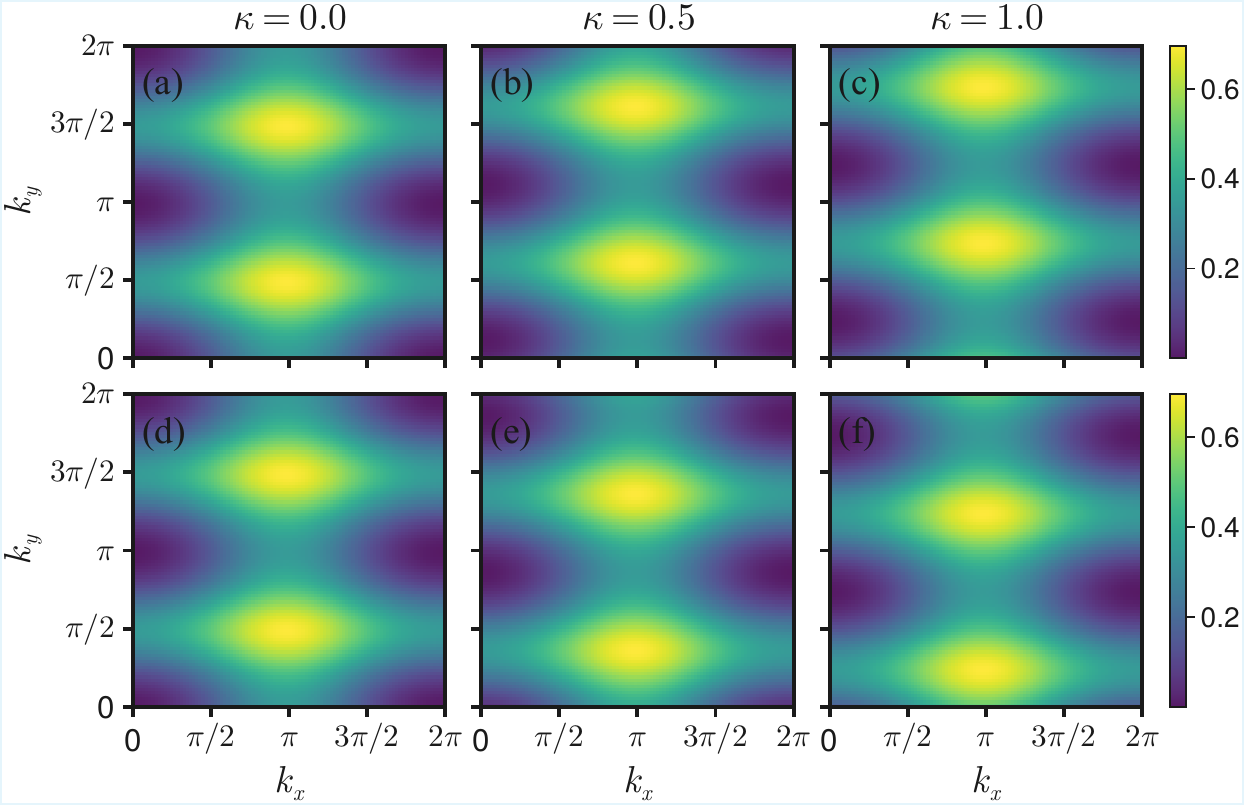}\hfill
  \caption{\textbf{}$\mathrm{Tr}(g^{RR})(=\Omega_{xy}^{RR})$ (upper row) and $\mathrm{Tr}(g^{LL})(=\Omega_{xy}^{LL})$ (lower row) for (a)(d) $\kappa=0$, (b)(e) $\kappa=0.5$ and (c)(f) $\kappa=1.0$.
  }
  \label{fig_SM_QG}
\end{figure}
In this section, we show the quantum geometry of the non-Hermitian ideal Chern band calculated with solely right or left eigenvectors and provide analytical explanations.

We numerically find both $g_{\mu\nu}^{XX}$ and $\Omega_{\mu\nu}^{XX}$ defined in Eq.~(5) and (6) in the main text are purely real for $X=R$ and $X=L$.
The ideal trace condition is satisfied, i.e.,
\begin{align}
\mathrm{Tr}(g^{XX})=\Omega^{XX}_{xy}.
\end{align}
In Fig.~\ref{fig_SM_QG}, we show $\mathrm{Tr}(g^{RR})$ [$\mathrm{Tr}(g^{LL})$] equal to $\Omega_{xy}^{RR}$ ($\Omega_{xy}^{LL}$) in the upper (lower) row for $\kappa=0.0,0.5,1.0$, respectively, from left to right.
We see the non-Hermitian deformation induces a shift of the quantum geometry pattern along positive (negative) direction along $k_y$, which we will explain analytically in the following. 

Again, we start from the continuum problem as in Sec.\ref{AWF}. 
For consistency with the numerical results, we use the Landau gauge $\bm{A}=(0,Bx,0)$ and obtain ladder operators
\begin{align}
a&=\frac{l_{B}}{\sqrt{2}\hbar}\bigg(\hbar\partial_{x}-i\hbar\partial_{y}-qBx\bigg),\\
a^{\dagger}&=\frac{l_{B}}{\sqrt{2}\hbar}\bigg(-\hbar\partial_{x}-i\hbar\partial_{y}-qBx\bigg)
\end{align}
for Eq.~\ref{HamLL}.
In the momentum space, the ladder operators take the form of  
\begin{align}
a&=\frac{l_{B}}{\sqrt{2}\hbar}\bigg(i\hbar k_{x}+\hbar k_{y}+iqB\partial_{k_x}\bigg),\\
a^{\dagger}&=\frac{l_{B}}{\sqrt{2}\hbar}\bigg(-i\hbar k_{x}+\hbar k_{y}+iqB\partial_{k_x}\bigg).
\end{align}
From Eq.~\ref{b1} and \ref{b2}, we obtain
\begin{align}
b&=\frac{l_{B}}{\sqrt{2}\hbar}\bigg(i\hbar k_{x}+\hbar k^R_{y}+iqB\partial_{k_x}\bigg),\\
b^{+}&=\frac{l_{B}}{\sqrt{2}\hbar}\bigg(-i\hbar k_{x}+\hbar k^L_{y}+iqB\partial_{k_x}\bigg)
\end{align}
where $k^R_y=k_y+q\gamma/\hbar,k^L_y=k_y-q\gamma/\hbar$.
Therefore, following the similar reasoning in Sec.~\ref{AWF}, the right and left eigenvectors on the lowest Landau level are $\psi^R(k_x,k_y)=\psi(k_x,k_y+q\gamma/\hbar),\psi^L(k_x,k_y)=\psi(k_x,k_y-q\gamma/\hbar)$ with $\psi(k_x,k_y)$ being the Hermitian Landau level wave function.
Utilizing the mapping between lowest Landau level wave functions in the continuum and flat band in the KM model even in the non-Hermitian regime proved in Sec.~\ref{AWF}, 
we arrive at $\psi^R_{KM}(k_x,k_y)=\psi_{KM}(k_x,k_y+\kappa(\pi/2)(1-\phi)),\psi^L_{KM}(k_x,k_y)=\psi_{KM}(k_x,k_y-\kappa(\pi/2)(1-\phi))$ for the lattice wave functions on the flat band,
which explains the spatial shift of the quantum geometry pattern in Fig.~\ref{fig_SM_QG} under the non-Hermitian deformation.

\section{Single-band projection in the non-Hermitian context}
In this section, we provide details of the single-band projection in the non-Hermitian context and show the validity of this approximation numerically.

For generality, we consider a general free-particle Hamiltonian with translational invariance
\begin{align}\label{Hf}
H_f&=\sum_{k,\alpha,\beta} a^\dagger_{k,\alpha}H^{\alpha,\beta}_f(k)a_{k,\beta},
\end{align}
where $k$ is the index of the lattice momentum, $\alpha,\beta$ are internal degrees of freedom within a unit cell and $H^{\alpha,\beta}_f(k)$ is the matrix element of the first-quantized Hamiltonian $H_f(k)$.
For a non-Hermitian system, $H_f(k)$ is a non-Hermitian matrix which can be diagonalized as
\begin{align}
H_f(k)=\sum_{n,k}\epsilon_{n,k}|u^R_n(k)\rangle \langle u^L_n(k)|,
\end{align}
where $\epsilon_{n,k}$ is the eigenenergy of band $n$ and $|u^R_n(k)\rangle$ and $|u^L_n(k)\rangle$ are the corresponding right and left eigenvectors satisfying $\langle u^L_n(k)|u^R_m(k)\rangle=\delta_{nm}$.
The second quantized Hamiltonian Eq.~\ref{Hf} can be written as 
\begin{align}
H_f&=\sum_{n,k} \epsilon_{n,k}c^+_{n,k}c_{n,k},
\end{align}
where 
\begin{align}
c^+_{n,k}&=\sum_{\alpha} [u^R_{n}(k)]_\alpha a^\dagger_{k,\alpha},\label{ac1}\\
c_{n,k}&=\sum_{\alpha} [u^L_{n}(k)]^*_\alpha a_{k,\alpha}, \label{ac2}
\end{align}
with $[u^R_{n}(k)]_\alpha$ ($[u^L_{n}(k)]_\alpha$) being the amplitude of the wave function $|u^R_n(k)\rangle$ ($|u^L_n(k)\rangle$) on the internal orbital $\alpha$.
Due to the biorthonormalization relation $\langle u^L_n(k)|u^R_m(k)\rangle=\delta_{nm}$, $c_{n,k}$ and $c^+_{n,k}$ keep the original statistics of $a_{k,\beta}$ and $a^\dagger_{k,\alpha}$, i.e., for bosons satisfying the commuting relations $[c_{n_1,k_1},c_{n_2,k_2}]=0,[c^+_{n_1,k_1},c^+_{n_2,k_2}]=0,[c_{n_1,k_1},c^+_{n_2,k_2}]=\delta_{n_1n_2}\delta_{k_1k_2}$ and 
for fermions satisfying the anti-commuting relations $\{c_{n_1,k_1},c_{n_2,k_2}\}=0,\{c^+_{n_1,k_1},c^+_{n_2,k_2}\}=0,\{c_{n_1,k_1},c^+_{n_2,k_2}\}=\delta_{n_1n_2}\delta_{k_1k_2}$.
However, $c_{n,k}$ and $c^+_{n,k}$ are not Hermitian conjugate of each other, thus we use $+$ instead of $\dagger$.
Therefore, $c_{n,k}$ and $c^+_{n,k}$ can be viewed as non-Hermitian generalization of creation and annihilation operators. 

Eq.~\ref{ac1} and \ref{ac2} represent a similarity transformation because of the biorthonormalization relation.
Therefore we can get the inverse transformation
\begin{align}
a^\dagger_{k,\alpha}&=\sum_{n} [u^L_{n}(k)]^*_\alpha c^+_{n,k},\\ 
a_{k,\alpha}&=\sum_{n} [u^R_{n}(k)]_\alpha c_{n,k}.
\end{align}
Then for a general interaction in momentum space
\begin{align}
V_I&=\sum_{k_1,\alpha_1,k_2,\alpha_2,k_3,\alpha_3,k_4,\alpha_4}U(k_1,\alpha_1,k_2,\alpha_2,k_3,\alpha_3,k_4,\alpha_4)a^\dagger_{k_1,\alpha_1}a^\dagger_{k_2,\alpha_2}a_{k_3,\alpha_3}a_{k_4,\alpha_4}\\
&=\sum_{k_1,n_1,k_2,n_2,k_3,n_3,k_4,n_4}\tilde{U}(k_1,n_1,k_2,n_2,k_3,n_3,k_4,n_4)c^+_{n_1,k_1}c^+_{n_2,k_2}c_{n_3,k_3}c_{n_4,k_4},
\end{align}
where 
\begin{align}
\tilde{U}(k_1,n_1,k_2,n_2,k_3,n_3,k_4,n_4)=\sum_{\alpha_1,\alpha_2,\alpha_3,\alpha_4}U(k_1,\alpha_1,k_2,\alpha_2,k_3,\alpha_3,k_4,\alpha_4)[u^L_{n_1}(k_1)]^*_{\alpha_1}[u^L_{n_2}(k_2)]^*_{\alpha_2}[u^R_{n_3}(k_3)]_{\alpha_3}[u^R_{n_4}(k_4)]_{\alpha_4}.
\end{align}
Numerically, we can diagonalize $V_I$ in the Fock basis associated with $c_{n,k}$ and $c^+_{n,k}$ like in the conventional Hermitian Fock basis (actually biorthogonal Fock basis here).
If we focus on a single band indexed by $n_0$ and assume the effect from other bands is negligible,
we arrive at
\begin{align}
V_I\approx  \sum_{k_1,k_2,k_3,k_4}\tilde{U}(k_1,n_0,k_2,n_0,k_3,n_0,k_4,n_0)c^+_{n_0,k_1}c^+_{n_0,k_2}c_{n_0,k_3}c_{n_0,k_4}.
\end{align}

\begin{figure}[h]
  \centering
  \includegraphics[width=1\textwidth]{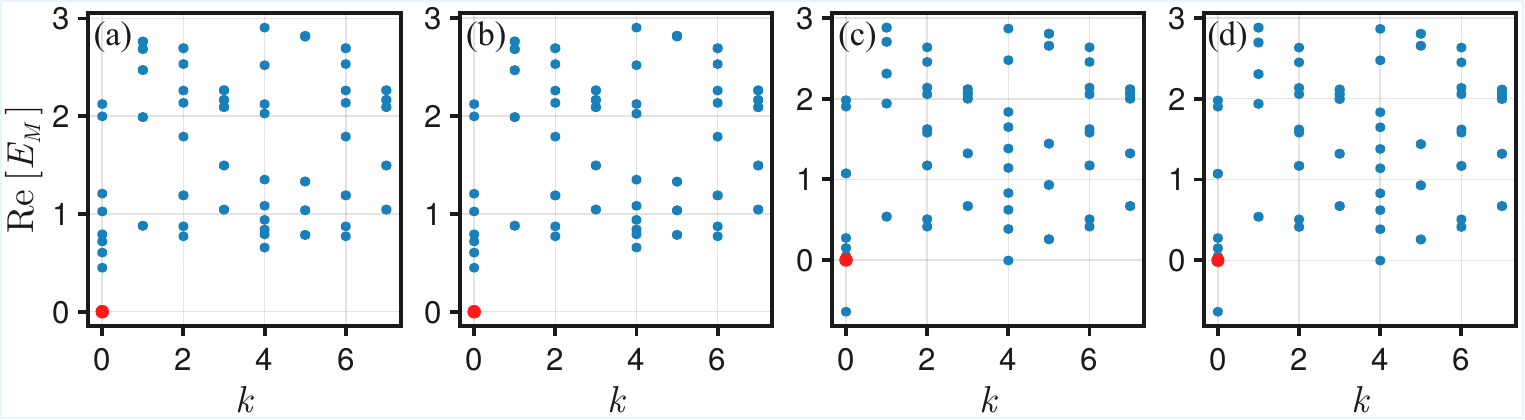}\hfill
  \caption{\textbf{}Real part of the many-body energy spectrum with and without the single-band projection for system size $L_x=4,L_y=4$ and flux $\phi=1/2$. (a) With single-band projection at $\kappa=0.5$. 
  (b) Without single-band projection at $\kappa=0.5$ and $U=0.001$.
  (c) With single-band projection at $\kappa=0.7$. 
  (d) Without single-band projection at $\kappa=0.7$ and $U=0.001$.
  For (b) and (d), the single-particle energy has been subtracted and the energy has been renormalized to the same scale as (a) and (c).
  }
  \label{fig_Validity}
\end{figure}

To validate the single-band projection for non-Hermitian systems, we calculate the many-body spectrum of Eq.~9 in the main text both with and without the projection, as shown in Fig.~\ref{fig_Validity}.
Fig.~\ref{fig_Validity}(a) and (b) shows the real part of the energy spectrum at $\kappa=0.5$, at which zero modes are ground states, and (a) is calculated under single-band projection and (b) without projection at interaction strength $U=0.001$.
We see the energy spectrum in (a) and (b) are the same.
Note that for (b) the kinetic energy $E_{kin}=n_bE_{flat}$ has been subtracted with $n_b$ being the particle number and $E_{flat}=-1$ the single-particle energy of the flat band, and the energy has been normalized to the same scale as (a) where the interaction strength is the only energy scale and chosen arbitrarily.
Similarly, Fig.~\ref{fig_Validity}(c) and (d) are also the same for $\kappa=0.7$, after the zero modes transition out of the ground state sector.
Therefore, we show the validity of the single-band projection in the non-Hermitian context.

\section{Spectral Flow}
In this section, we exhibit the spectral flow to show the existence of a finite gap in the regime with fractional Chern insulators (FCIs) as ground states, and the gapless nature of the system after FCIs transition out of the ground state sector. 

\begin{figure}[h]
  \centering
  \includegraphics[width=1\textwidth]{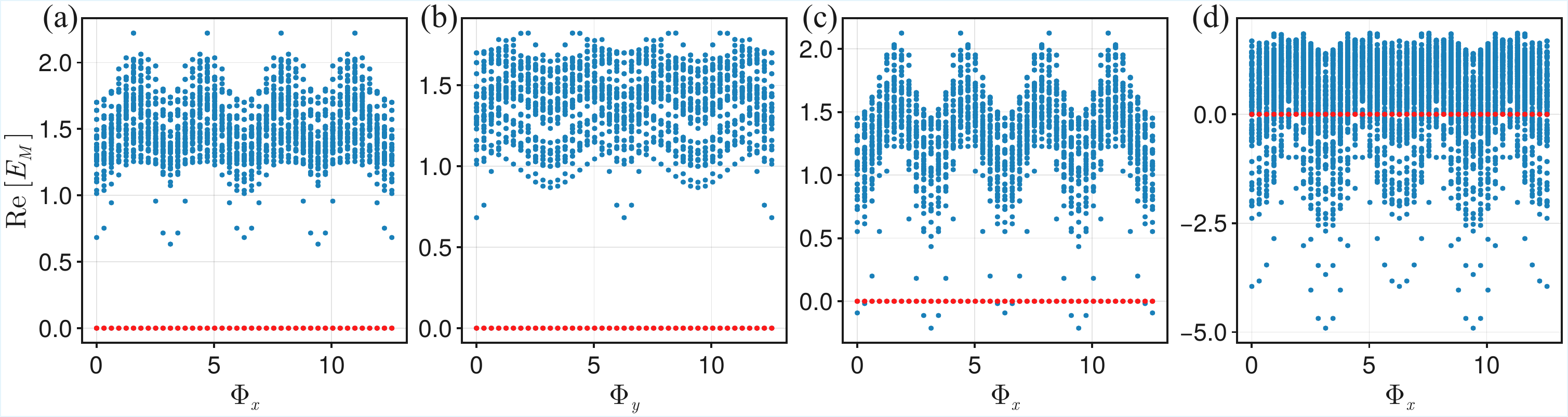}\hfill
  \caption{\textbf{}(a)-(b) Spectral flow at $\kappa=0.6$ about (a) $\Phi_x$ and (b) $\Phi_y$.
  (a)-(b) Spectral flow about $\Phi_x$ at (c) $\kappa=0.67$ and (d) $\kappa=0.9$.
  The system size is $L_x=6,L_y=4$ and flux is $\phi=1/2$.
  }
  \label{fig_SpectralFlow}
\end{figure}

In Fig.~\ref{fig_SpectralFlow}, We show the spectral flow under twisted boundary phase $\Phi_x$ ($\Phi_y$) along $x$ ($y$) direction.
We show the spectral flow at $\kappa=0.6$ about $\Phi_x$ and $\Phi_y$ in Fig.~\ref{fig_SpectralFlow}(a) and (b), respectively, where we see a persistent finite gap between the FCIs (indicated by red color) and excited states.
At $\kappa=0.67$, close to the transition point, as shown in Fig.~\ref{fig_SpectralFlow}(c), we see the single state with negative energy under no twisted boundary phase can float above zero energy during the spectral flow and become indistinguishable from the dense excited states at some points, implying the system is gapless.
When we further increase $\kappa$ to $0.9$, as shown in Fig.~\ref{fig_SpectralFlow}(d), more states drop below zero energy and we cannot identify an isolated manifold of states in the spectral flow pattern, further indicating the gapless nature of the system.

\section{Energy spectrum in the complex energy plane and under complex interaction $U$}\label{ComplexU}

In this section, we exhibit the many-body energy spectrum in the complex energy plane and demonstrate that the FCIs can be engineered as the longest-lived states by complex interaction strength $U$.

\begin{figure}[h]
  \centering
  \includegraphics[width=1\textwidth]{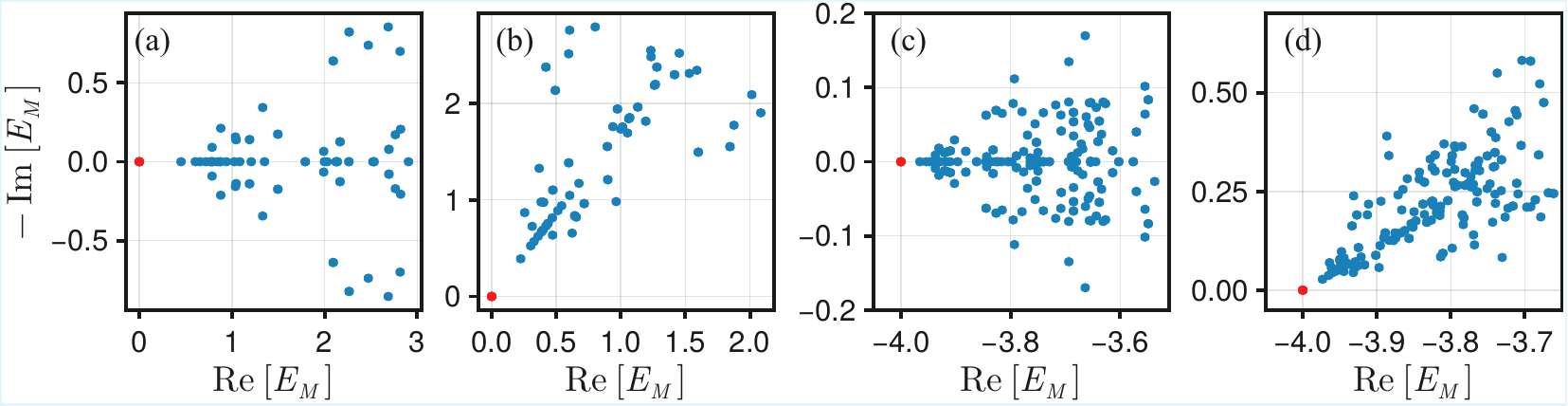}\hfill
  \caption{\textbf{} Many-body energy spectrum in the complex plane for $\kappa=0.5,L_x=L_y=4, n_b=4, \phi=1/2$. The energies are calculated under single-band projection with $|U|=1$ for (a),(b) while without the projection with $|U|=0.1$ for (c),(d).  
  For (a),(c), $U$ is real, i.e., $\theta=0$ while for (b),(d), $\theta=-\pi/3$.
  }
  \label{fig_ImagU}
\end{figure}

For the interacting non-Hermitian KM model in the topologically ordered regime, at real $U$ under single-band projection, the FCIs are steady states as the energy is purely real, as indicated by red color in Fig.~\ref{fig_ImagU}(a).
On the other hand, the energy spectrum is symmetric about the real axis, indicating excited states with positive $\mathrm{Im}\:[E_M]$ will amplify with time.   
If we choose complex $U=|U|e^{i\theta}$, which can be realized by particle loss in realistic experiments~\cite{Yoshida2019SP}, the energy spectrum is rotated by $\theta$ in the complex energy plane about $E_M=0$, the energy of the FCIs.
Therefore, as long as the pattern of the complex energy spectrum can be enveloped within an acute angle with vertex at $E_M=0$, there always exists a range of $\theta$ such that all the excited states have positive real part and negative imaginary part of the energy, as shown in Fig.~\ref{fig_ImagU}(b).
In other words, the FCIs have the longest lifetime, justifying the possible experimental realization in principle.
Without the single-band projection, the result still holds, as shown in Fig.~\ref{fig_ImagU}(c) and (d).

\section{Entanglement Spectrum}
In this section, we show more data on the entanglement spectrum to confirm the zero modes are FCIs.

\begin{figure}[h]
  \centering
  \includegraphics[width=0.8\textwidth]{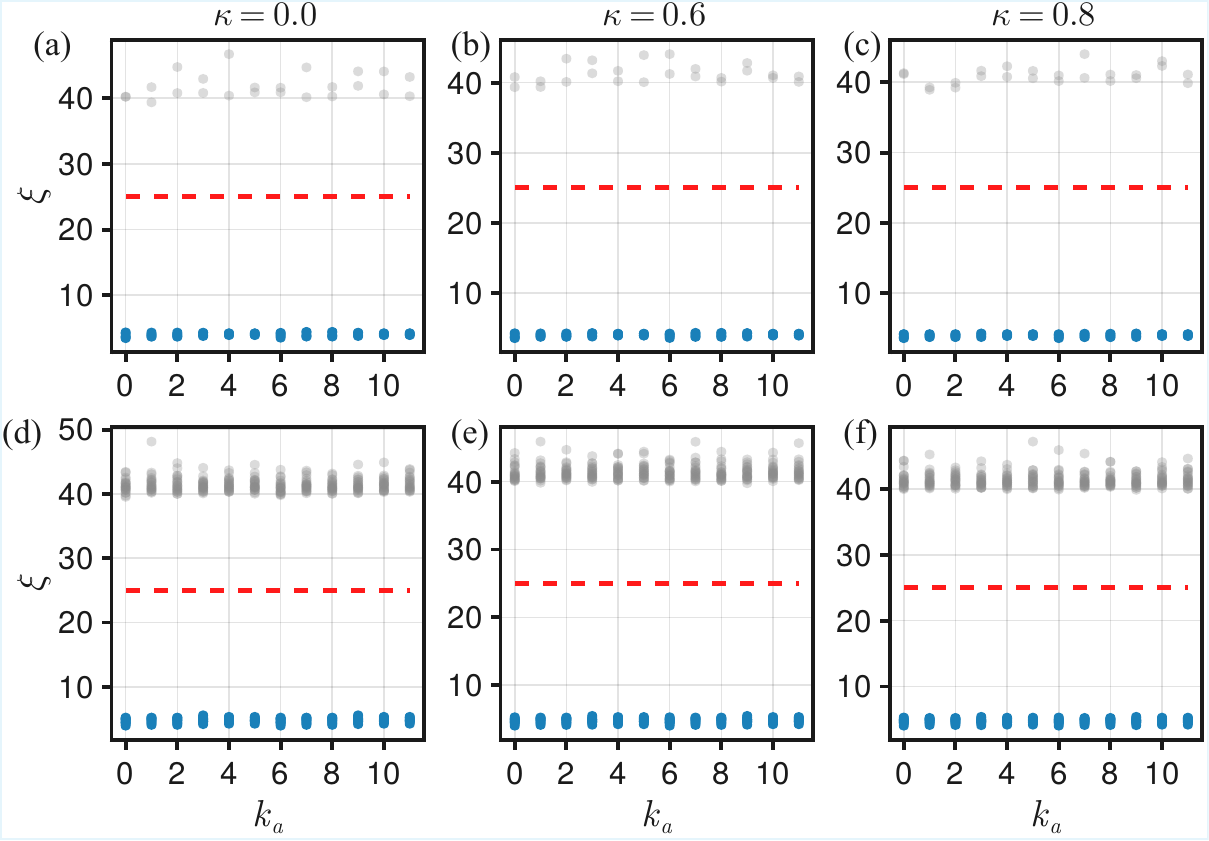}\hfill
  \caption{\textbf{} Entanglement spectrum calculated using the right eigenvectors of the zero modes for $L_x=6,L_y=4,\phi=1/2$.
  $n_A=2$ for (a)-(c) and $n_A=3$ for (d)-(f).
  }
  \label{fig_ESw}
\end{figure}

We calculate the entanglement spectrum using the right eigenvectors of the zero modes,
i.e., $X=X^{\prime}=R$ for $\rho_A=\mathrm{Tr}_B(\sum_i|\Psi^X_i\rangle\langle\Psi^{X^\prime}_i|/2)$,
for system size $L_x=6,L_y=4$ as shown in Fig.~\ref{fig_ESw}. 
We show results for the Hermitian case ($\kappa=0$) in the left column, the non-Hermitian FCI phase ($\kappa=0.6$) in the middle column and the phase with negative energy ($\kappa=0.8$) in the right column.
In all the subfigures, we see a large gap indicated by the red dashed line. 
The number of states below the gap is 54 for the first row with $n_A=2$ and 112 for the second row with $n_A=3$, consistent with the generalized Pauli principle.
Therefore, the entanglement spectrum verifies that the exact zero modes are FCIs,
even in the presence of negative energy states.
Note that $\xi$ of the states above the gap is infinite in principle and arises at finite value in our plot due to the machine precision. 

\section{Simple interpretation of negative energy states}
In this section, we give a simple interpretation for why negative energy states in Fig.~3(d) in the main text can emerge under a positive definite interaction.

Our total interacting Hamiltonian is 
\begin{equation}\label{Int2}
H(\kappa)=H_0(\kappa)+\sum_j \frac{U}{2}a^\dagger_ja^\dagger_ja_ja_j,
\end{equation}
where the first term is the single-particle Kapit-Mueller model and the second is the interaction.
According to Weyl inequality, the lowest eigenvalue of the sum of two Hermitian operators $A$ and $B$, is greater than or equal to the sum of the lowest eigenvalues of $A$ and $B$, i.e, $\lambda_{A+B}^{\mathrm{min}}\geq \lambda_{A}^{\mathrm{min}}+\lambda_{B}^{\mathrm{min}}$.  
Therefore, in the Hermitian limit, the FCIs, which minimize both the kinetic and interaction terms, are always the ground states.

On the other hand, Weyl equality is only in the Hermitian regime, and we can find a counterexample even when only one of the two operators is non-Hermitian and has real spectra. 
Given two simple $2\times 2$ matrices:
\[
A = \begin{pmatrix}
-8 & -9 \\
9 & 10
\end{pmatrix},
\quad
B = \begin{pmatrix}
1 & 0 \\
0 & 2
\end{pmatrix},
\]
the eigenvalues of $A$ are $\lambda_A=1,1$ and those of $B$ are $\lambda_B=1,2$.
Although eigenvalues of $A$ and $B$ are all positive, the eigenvalues of 
\[
S=A+B = \begin{pmatrix}
-7 & -9 \\
9 & 12
\end{pmatrix}
\]
are $\lambda_S \approx 5.541,-0.541 $, with a negative one.
Therefore, minimizing kinetic energy and interaction ennergy, respectively, do not necessarily lead to the lowest total energy in the non-Hermitian regime. 
This explains why states in Fig.~3(b)(d) under large non-Hermiticity can counterintuitively acquire negative energy with a positive definite interaction, which is attributed to the distinct feature of non-Hermitian operators. 

\section{Signatures of superfluid}
In this section, we show evidence of superfluid for the negative energy states competing with FCIs by calculating the superfluid density.

\begin{figure}[h]
  \centering
  \includegraphics[width=0.5\textwidth]{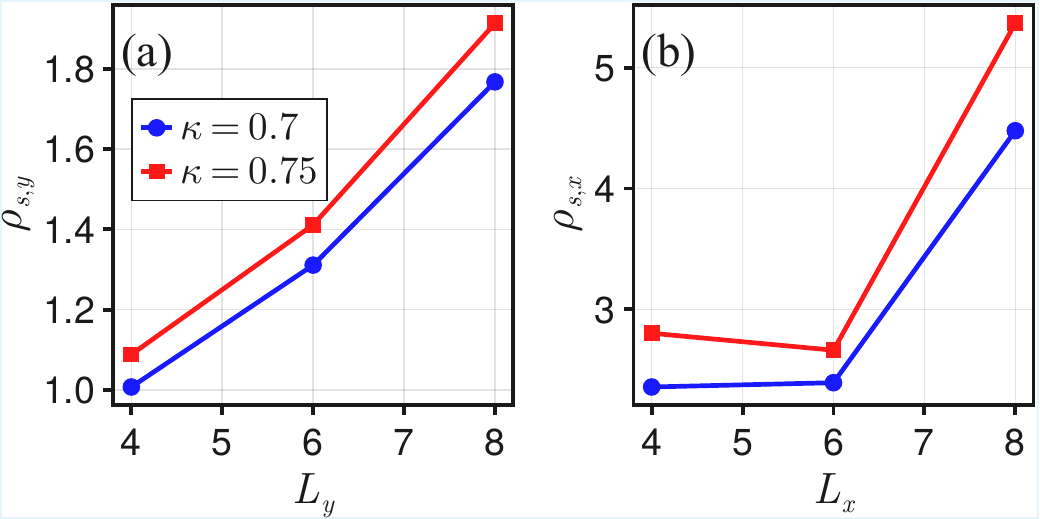}\hfill
  \caption{\textbf{}Superfluid density (a) $\rho_{s,y}$ as a function of $L_y$ for $L_x=4$ and $\rho_{s,x}$ as a function of $L_x$ for $L_y=4$. 
  }
  \label{fig_SuperDen}
\end{figure}

We calculate the superfluid density by the second derivative of the many-body state energy with respect to the twisted boundary phase $\Phi_x$ ($\Phi_y$) close to $\Phi_x=0$ ($\Phi_y=0$)~\cite{Fisher1973PRA,Paramekanti1998PRB}
\begin{align}
\rho_{s,y}=\frac{L_y}{L_x}\frac{\partial^2E_M}{\partial \Phi_y^2}\Big|_{\Phi_y=0},\\
\rho_{s,x}=\frac{L_x}{L_y}\frac{\partial^2E_M}{\partial \Phi_x^2}\Big|_{\Phi_x=0},
\end{align}
where $\rho_{s,x}$ and $\rho_{s,y}$ are superfluid density along $x$ and $y$ directions, respectively.
We show $\rho_{s,x}$ and $\rho_{s,y}$ of the lowest energy state at $\kappa=0.7$ and $0.75$ in Fig.~\ref{fig_SuperDen}, for which the energy is negative.
We see as the system size $L_x$ and $L_y$ increase, $\rho_{s,x}$ and $\rho_{s,y}$ grow in overall,
implying the negative energy states host nonzero superfluid density.

\section{Beyond the single-band projection}
In this section, we demonstrate the many-body energy spectrum under larger interaction strength without single-band projection.

\begin{figure}[h]
  \centering
  \includegraphics[width=0.8\textwidth]{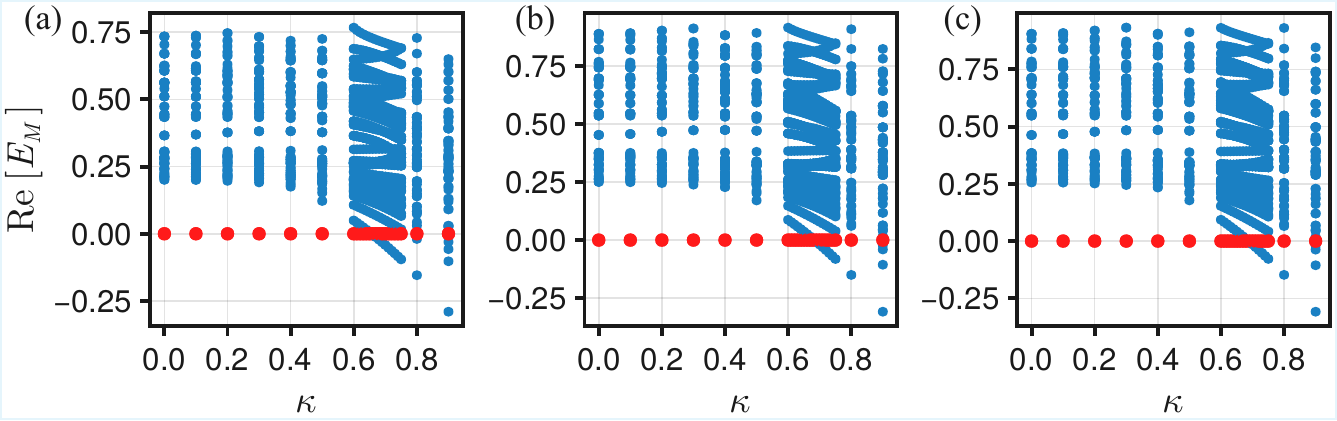}\hfill
  \caption{\textbf{}Many-body energy spectrum without the single-band projection as a function of $\kappa$ under (a) $U=1$, (b)$U=10$ and (c) $U=100$ at $L_x=4,L_y=4,\phi=1/2$.
  }
  \label{fig_Utran}
\end{figure}

We show the many-body energy spectrum without the single-band projection as a function of $\kappa$ under $U=1,10$ and $100$ in Fig.~\ref{fig_Utran}(a),(b) and (c), respectively.
For all the cases, we see FCIs as the ground states below a critical $\kappa$ and transitions into a regime with negative energy states for larger $\kappa$.
$U=100$ shows the physics in the hard-core limit, on the opposite side of the single-band projection.
Therefore our finding is general for the whole range of interaction strength. 

\begin{figure}[h]
  \centering
  \includegraphics[width=1\textwidth]{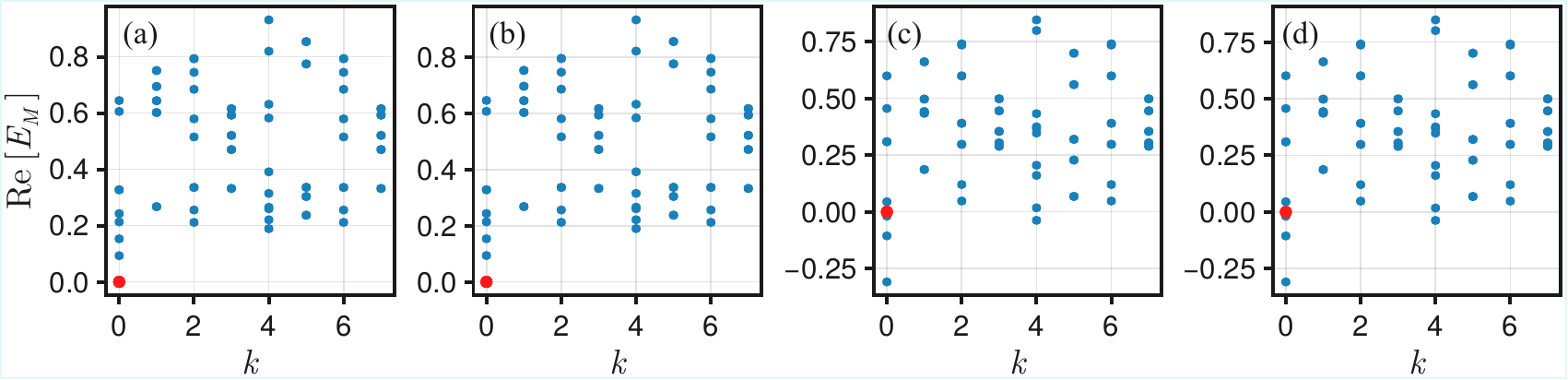}\hfill
  \caption{\textbf{}Real part of the many-body energy spectrum at $L_x=L_y=4,\phi=1/2$. (a) $U=100,\kappa=0.6$,
  (b) $U=1000,\kappa=0.6$, (c) $U=100,\kappa=0.9$,
  (d) $U=1000,\kappa=0.9$.
  }
  \label{fig_hardcore}
\end{figure}

To show $U=100$ is in the hard-core limit, we compare the many-body energy spectrum of $U=100$ and $U=1000$ as shown in Fig.~\ref{fig_hardcore}.
We cannot tell the difference between (a) with $U=100$ and (b) with $U=1000$ at $\kappa=0.6$ and
similarly (c) and (d) at $\kappa=0.9$, indicating $U=100$ is large enough to show the physics of hard-core bosons.

\section{Illustration of the breakdown of variational principle in non-Hermitian systems}
In this section, we will give a detailed explanation on the breakdown of the variational principle in non-Hermitian systems and its effect on the critical non-Hermiticity strength $\kappa_c$ under different interacting strength.

Let us recall the variational principle in Hermitian quantum mechanics first.
For a Hermitian Hamiltonian $H$ acting on a Hilbert space $\mathcal{H}$, the expectation value $\langle \psi|H|\psi\rangle\ge E_G$ for any state $|\psi\rangle$ in $\mathcal{H}$ and $E_G$ is the ground state energy of $H$, and the equality holds when $|\psi\rangle$ takes the ground state $|\psi_G\rangle$.
Given a subspace $\mathcal{H}_1\subseteq \mathcal{H}$, we can write the matrix representation of the Hamiltonian $H$ as
\[
H = \begin{pmatrix}
H_1 & H_{12} \\
H_{21} & H_2
\end{pmatrix},
\]
where $H_1,H_2,H_{12},H_{21}$ are matrices and $H_1$ is the matrix representation in subspace $\mathcal{H}_1$.
The variational principle can be formulated as $E_1\ge E_G$ with $E_1$ being the lowest eigenvalue of $H_1$, because $E_1=\langle \psi_1|H|\psi_1\rangle\ge E_G$ with the eigenvector $|\psi_1\rangle$ of $H_1$ being a vector in both $\mathcal{H}_1$ and $\mathcal{H}$. 

In our interacting model, as shown in Fig.~3(d) in the main text, the lowest real part of the energy of the states other than the zero-energy FCIs decreases as the non-Hermiticity strength $\kappa$ increases, and drops to below zero after a critical value.  
The single-band projection corresponds to the variation within a sub Hilbert space, thus the lowest eigenenergy under projection cannot be lower than the ground state energy obtained by diagonalizing the complete Hamiltonian without single-band projection, i.e., $E_G^{\mathrm{Proj}}(\kappa)\geq E_G^{\mathrm{Full}}(\kappa)$, if the variational principle still applies.  
If that is the case, by diagonalizing the complete Hamiltonian, the lowest real part of the energy other than the zero modes would be zero or negative at the critical value under the single-band projection $\kappa_{c,s}$, i.e., $E_G^{\mathrm{Proj}}(\kappa_{c,s})\geq E_G^{\mathrm{Full}}(\kappa_{c,s})$. 
Therefore, as $E_G(\kappa_1)\geq E_G(\kappa_2)$ for $\kappa_1\leq \kappa_2$ obtained by numerics, the critical value for the full interacting model $\kappa_{c,f}$ is not larger than $\kappa_{c,s}$.
However, our numerics demonstrate that $\kappa_c$ increases as we discard the single-band projection,
evidencing the breakdown of the variational principle in non-Hermitian systems.

We show the breakdown by an example of a $2\times 2$ matrix:
\[
H = \begin{pmatrix}
0 & -1 \\
1 & 2
\end{pmatrix},
\]
where $H_1=0$ can be considered as the matrix representation in a restricted Hilbert space of one dimension, with lowest eigenvalue $E_1=0$.
The eigenvalues of $H$ are $1,1$, i.e., $E_G=1$, obviously violating the variational principle $E_1\ge E_G$.

\section{$\phi=1/3$ case}
In this section, we show numerical results at flux $\phi=1/3$ to illuminate the generality of our findings.

\begin{figure}[h]
  \centering
  \includegraphics[width=1\textwidth]{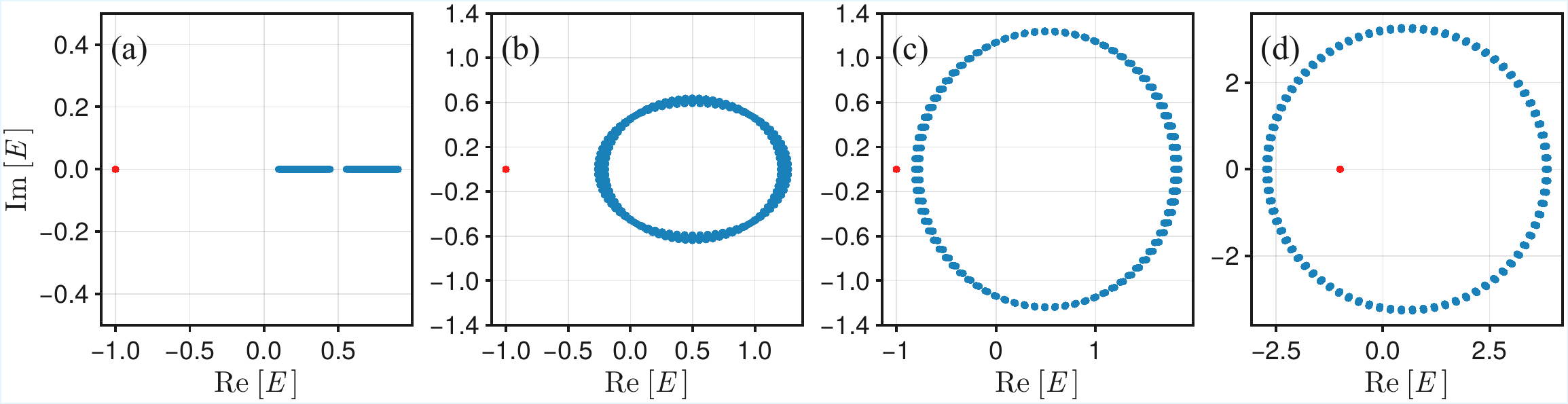}\hfill
  \caption{\textbf{}Single-particle energy in the complex energy plane of the non-Hermitian KM model at $\phi=1/3$ for (a) $\kappa=0$ (b) $\kappa=1$ (c) $\kappa=1.4$ (d) $\kappa=2$.
  }
  \label{fig_onethird_sing}
\end{figure}

In Fig.~\ref{fig_onethird_sing}, we show the complex energy of the non-Hermitian KM model in Eq.~1 of the main text with $\kappa=0,1,1.4,2$ from left to right.
The same as the $\phi=1/2$ case, we see the flat band also remains exactly real and flat under the non-Hermitian deformation. 
The flat band energy is pinned at $E=-1$ and the loop formed by excited states enlarges as $\kappa$ increases.
After the gap closing, the flat band tunnels into the loop while remain intact.

\begin{figure}[h]
  \centering
  \includegraphics[width=0.6\textwidth]{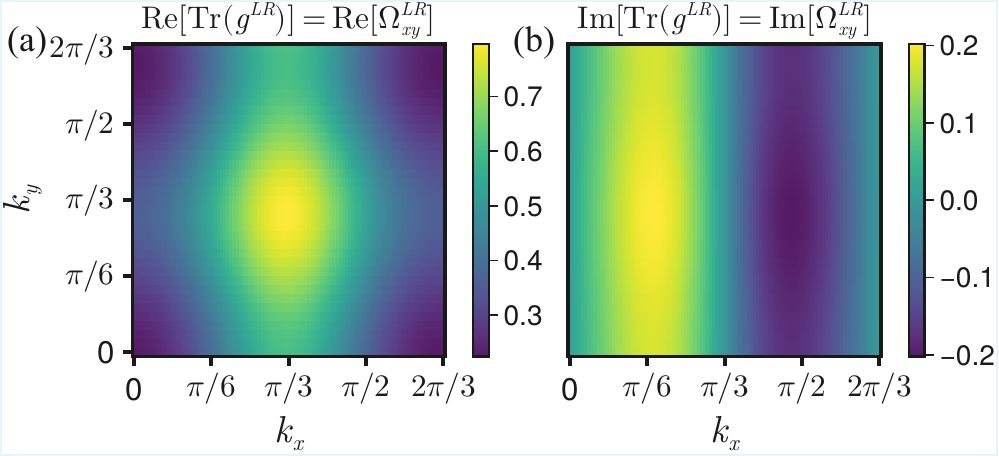}\hfill
  \caption{\textbf{}(a) Real and (b) imaginary part of the quantum geometry at $\phi=1/3$ of one-third of the magnetic Brillouin zone. 
  }
  \label{fig_SM_QG_third}
\end{figure}

The real and imaginary parts of the quantum geometry at $\phi=1/3$ are shown in Fig.~\ref{fig_SM_QG_third}(a) and (b), respectively.
The quantum geometry at $\phi=1/3$ also satisfies the generalized ideal condition in Eq.~(7)(8) in the main text.

\begin{figure}[h]
  \centering
  \includegraphics[width=1.0\textwidth]{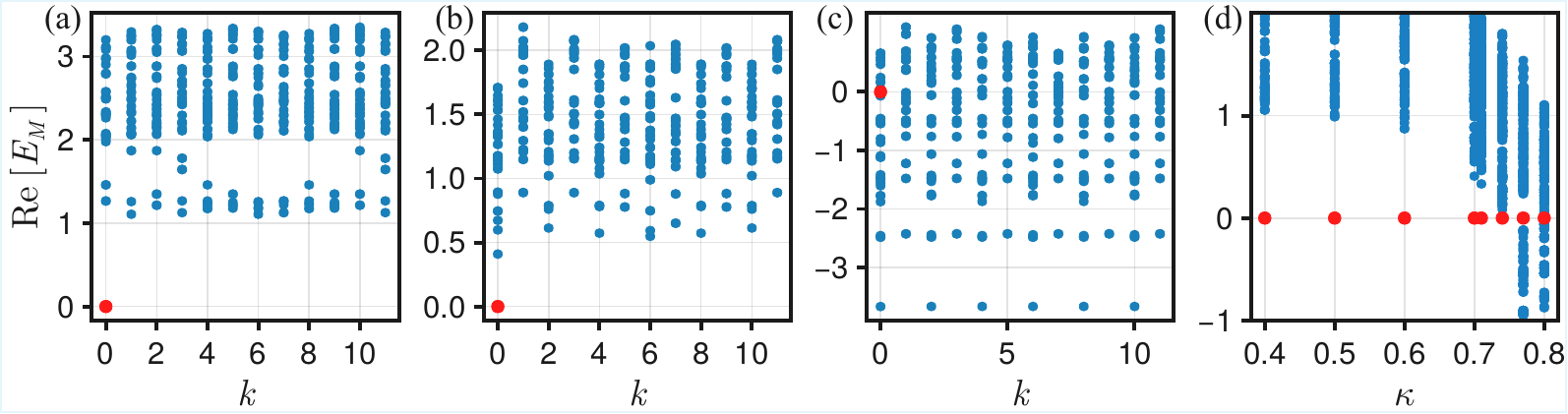}\hfill
  \caption{\textbf{} Real part of the many-body energy spectrum for $L_x=6,L_y=6,\phi=1/3$: (a) $\kappa=0$ (b) $\kappa=0.7$ (c) $\kappa=0.8$ (d) with respect to $\kappa$.
  }
  \label{fig_SM_onethird_many}
\end{figure}

In Fig.~\ref{fig_SM_onethird_many}, we show the real part of the many-body energy spectrum at $\kappa=0.0,0.7$ and $0.8$ in (a), (b) and (c), respectively.
We see zero-energy modes as the ground states at $\kappa=0.7$, while at $\kappa=0.8$, negative energy states emerge.
The transition is shown in Fig.~\ref{fig_SM_onethird_many}(c) by the energy spectrum as a function of $\kappa$.

\end{document}